\documentclass[11pt]{article}
\usepackage{array,subfigure,setspace}

\newtheorem{theorem}{Theorem}[section]

\newtheorem{prop}{Proposition}[section]
\newtheorem{lemma}{Lemma}[section]

\usepackage{amsfonts}
\usepackage{graphicx}
\usepackage{color,times}

\usepackage[authoryear]{natbib}
\renewcommand\cite{\citet}

\def\E{\mathbb E}
\def\noi{\noindent}
\def\bbN{\mathbb N}
\def\bbZ{\mathbb Z}
\def\wtilde{\widetilde}
\def\P{{\mathbb P}}
\def\bbR{{\mathbb R}}
\def\esssup{{\rm esssup}}
\def\what{\widehat}
\def\refeq#1{{(\ref{e:#1})}}

\title{On the estimation of the extremal index  based on scaling and resampling}

\begin{document}




\date{February 12, 2009}

\author{Kamal Hamidieh\\ {\it Department of Statistics, Rice University}\\ \ \ \  Stilian Stoev\ \  and\  \  George Michailidis\\
 {\it Department of Statistics,  The University of Michigan}}

\maketitle

\begin{abstract}
The extremal index parameter $\theta$ characterizes the degree of
local dependence in the extremes of a stationary time series and has important applications
in a number of areas, such as hydrology, telecommunications, finance and environmental studies.
In this study, a novel estimator for $\theta$ based on the asymptotic scaling of block--maxima and
resampling is introduced. It is shown to be consistent and asymptotically normal for a
large class of $m-$dependent time series. Further, a procedure for the automatic selection of its tuning
parameter is developed and different types of confidence intervals that prove useful in practice
proposed. The performance of the estimator is examined through simulations, which show its highly
competitive behavior. Finally, the estimator is applied to three real data sets of daily crude oil
prices, daily returns of the S\&P 500 stock index, and high--frequency, intra--day traded volumes of a stock.
These applications demonstrate additional diagnostic features of statistical plots based on the new estimator.

\vspace{9pt}
\par
\end{abstract}

\fontsize{10.95}{14pt plus.8pt minus .6pt}\selectfont

\section{Introduction}

\begin{figure}[t!]
\begin{center}
\subfigure[Negative Log--returns of WTI Oil prices]{\includegraphics[width=2.5in]{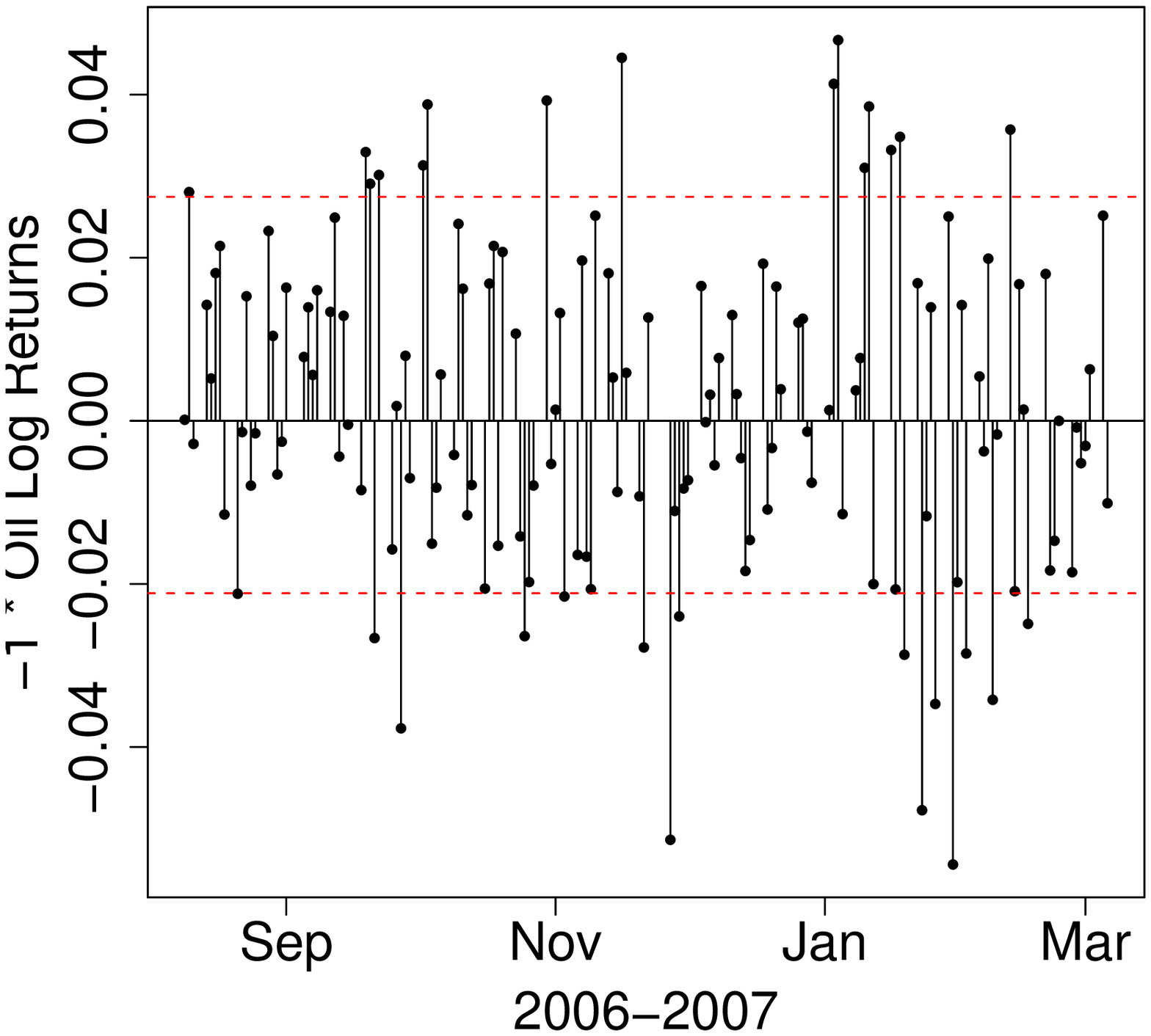}
\label{fig:portionofoilreturns}}
\subfigure[High--frequency Traded Volume]{\includegraphics[width=2.5in]{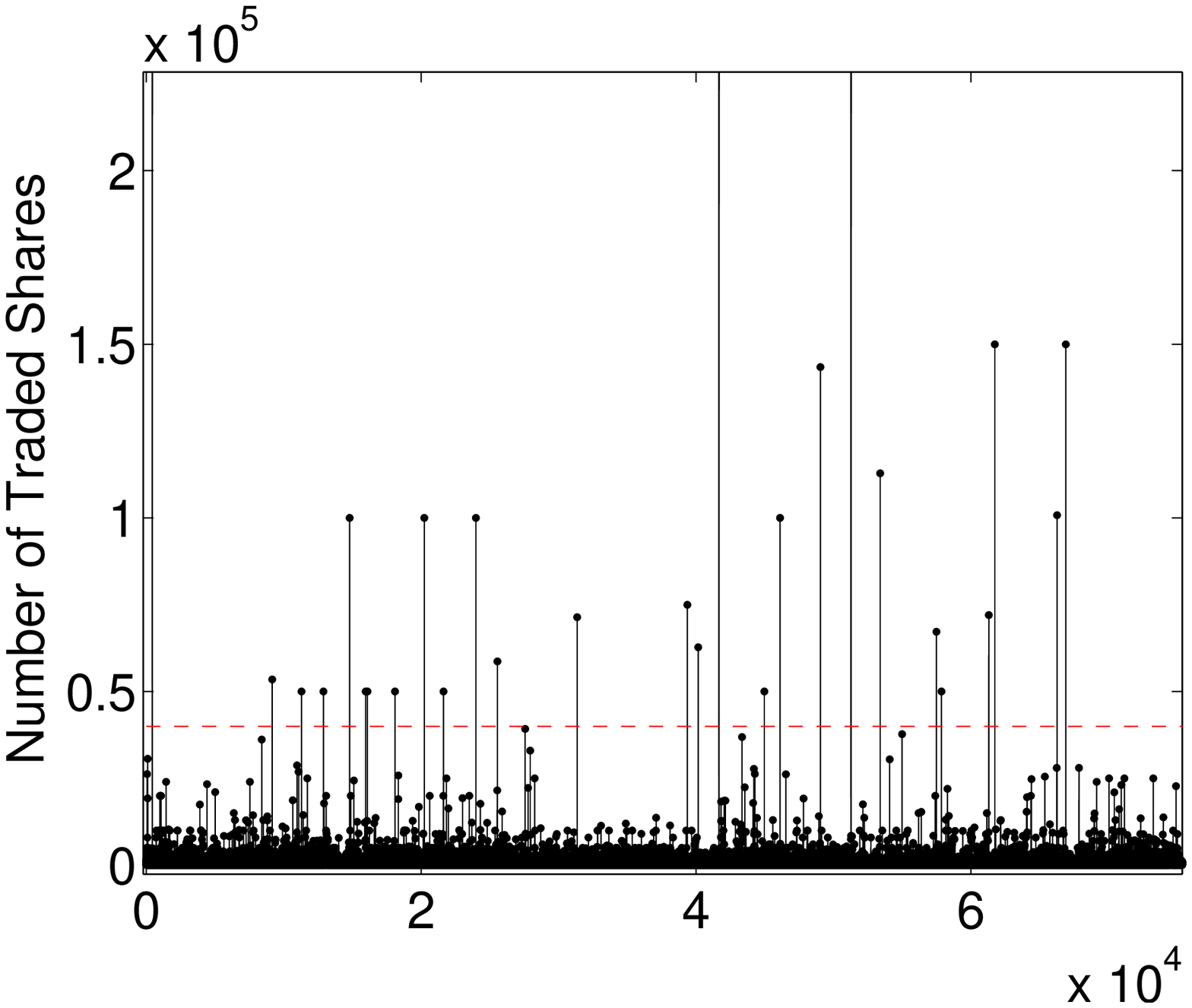}\label{fig:volume}}
{\caption{\label{fig1}  \footnotesize {\it Left plot:} Negative log--returns of daily Oil prices.  The upper and lower dashed lines
correspond to the 0.90 and 0.10 quantiles of the data respectively.  {\it Right plot:}
 High--frequency traded volumes (in numbers of shares per transaction) of the Intel stock in November 16, 2005.
 Observe the clustering of extremes, particularly evident in the extreme price drops or 'losses' (above the horizontal dotted line)
 for the Oil data.  The trades of the Intel stock with extremely large volumes also exhibit substantial clustering.
}}
\end{center}
\end{figure}

Advances in computer technology have enabled the collection by research organizations and
businesses of large time series data sets. These data sets are primarily characterized by the fine
granularity ({\em high frequency}) of the time intervals at which the observations are collected; for example,
Internet traffic is sampled at millisecond intervals, while stock trades at every second. Such time
series data are characterized by the presence of long range dependence (the autocorrelation
function decays at a polynomial rate) and the heavy tailed nature of the marginal distribution
(see, e.g.\ \cite{finkenstadt:rootzen:2004}). In many cases, another phenomenon can be observed, namely the presence of {\em clustering}
of very large or very small values ({\em extremes}) of the data (see e.g.\ Figure \ref{fig1}).
For example, in Internet traces this is the result of bursty arrivals, while in data on returns of a financial asset this is primarily due
to the arrival of an external market shock.

The daily log-returns of the spot price of West Texas Intermediate crude oil are shown for the period September 2006 --
March 2007 in
Figure \ref{fig:portionofoilreturns}. A pronounced temporal clustering of the extreme values can be seen, indicating the
presence of local dependence in the extremes. Figure \ref{fig:volume} also demonstrates the substantial clustering
of the extremely large traded volumes in the high--frequency data set of all intra--day trading activity of the Intel stock, for example.
Such clustering behavior is of interest to subject matter experts and it has important implications in
practice, since it concerns large consecutive changes associated with large financial 'losses' or 'gains'.
Therefore, quantifying the nature of the dependence structure as well as the duration of extreme events
becomes an essential part of the understanding of these time series data.

The extremal index $\theta$ is the main parameter that describes and quantifies the clustering
characteristics of the extreme values in many stationary time series. Its formal definition is given next.
Let $X = \{X_k \}_{k\in\bbZ}$ be a strictly stationary time series.
Define the following quantities
$$
 M_n := \max_{1\le k \le n} X_k \ \ \ \ \mbox { and }\ \ \ \ M_n^{iid} := \max_{1\le k \le n} \wtilde X_k,
$$
where the $\wtilde X_k$'s are independent and identically distributed (iid) random variables with the same
distribution as the $X_k$'s.
Formally, the time series $X$ is said to have an {\it  extremal index} $\theta$, if for some norming
sequences $c_n>0$ and $d_n$, we have
\begin{equation}\label{e:M_n-iid-M_n}
 \P\{c_n^{-1} (M_n^{iid} - d_n) \le x\} \stackrel{w}{\longrightarrow} H(x)\ \ \ \mbox{ and }\ \ \
 \P\{c_n^{-1} (M_n - d_n) \le x\} \stackrel{w}{\longrightarrow} H^{\theta}(x),
\end{equation}
where $H(\cdot)$ is a non--degenerate extreme value distribution (see e.g.\ p.\ 417 in
 \cite{embrecht:kluppelberg:mikosch:1997}).

An informal interpretation of
$\theta$ is given in \cite{leadbetter:lindgren:rootzen:1983}, namely
$\theta\approx$(mean cluster size)$^{-1}$. For example, for the crude oil log-returns discussed in
Section \ref{s:data-analysis}, the extremal
index is estimated to be around 0.6, which means that on the average, two large size 'losses'
or 'gains' are recorded in a relatively short time span.
The modeling and analysis of rare events (extremes) has been an active area in probability and statistics
(see e.g. \cite{embrecht:kluppelberg:mikosch:1997},
\cite{beirlant:goegebeur:segers:teugels:2004}).  In the context of extremes, the study and the estimation
of the extremal index $\theta$, plays an important role.

In this paper, we focus on the non--degenerate case when the extremal index $\theta$ is positive.  Observe that in this
case the same normalization and centering sequences for the partial maxima $M_n$ and $M_n^{iid}$ above
yield non--degenerate limit distributions.
The extremal index takes values in the interval $[0,1]$; a value close to 0 indicates
a very strong short range extremal dependence, while a value close to 1 a rather weak
dependence. In fact, for iid  $X_k$'s, by (\ref{e:M_n-iid-M_n}), we have $\theta = 1$.
The extremal index, however, characterizes only the dependence of the extremes in the time
series data and thus the data may still exhibit strong dependence, even though $\theta\approx 1$.
The case of $\theta=0$ is considered to be a pathological one.

Theoretical properties of the extremal index have been studied fairly extensively;
(\cite{obrien:1987}, \cite{hsing:husler:leadbetter:1988},
and references therein).
The problem of estimating $\theta$ has also received some attention in the literature:
\cite{hsing:1993}, \cite{smith:weissman:1994}, \cite{weissman:novak:1998} and \cite{ferro-segers}.
Applications of the extremal index in various scientific areas include its incorporation in
calculations of the Value-at-Risk measure (\cite{longin:2000} and Kl\"uppelberg in \cite{finkenstadt:rootzen:2004}),
in the study of the Nasdaq and S\&P 500 indices (\cite{galbraith:zernov:2006}) and in the study
of GARCH processes (\cite{laurini:2004}).  The estimation of the extremal index $\theta$ is an important practical problem
with rapidly expanding areas of application to finance, insurance, hydrology and telecommunications, to name a few
(for more details, see e.g.\ \cite{embrecht:kluppelberg:mikosch:1997} and \cite{finkenstadt:rootzen:2004}).

Most previous estimators of  $\theta$ exploit its connection to the point process of exceedances.
In this study, we introduce a new method for estimating $\theta$ based on the asymptotic scaling properties
of block--maxima and resampling. Specifically, let $X_1,\ldots, X_n$ be a data sample from a heavy--tailed
time series with positive extremal index $\theta$.
The maximum values of the data calculated over blocks of size $m$, scale at a rate $m^{1/\alpha}$,
where $\alpha>0$ denotes the {\em tail index} of the marginal distribution of the data. Further, the normalized
limit of the block maxima is proportional to $\theta^{1/\alpha} \sigma$, where $\sigma:= c_X^{1/\alpha}>0$ is an asymptotic
scale coefficient of the $X_k$'s. Thus, by examining a sequence of growing, dyadic block
 sizes $m=2^j,~1\leq j\leq \lfloor \log_2 n \rfloor, ~j\in\mathbb{N}$, and subsequently estimating the mean
 of logarithms of block--maxima one obtains
 {\em estimating equations} involving both the tail index $\alpha$ and the parameter $\theta^{1/\alpha}\sigma$.
 In these equations, the scale $\sigma$ and the extremal index $\theta$ are, however, coupled. In principle, $\theta$ can be calculated by solving an appropriate nonlinear equation, but the resulting estimate proves to be too variable. Hence, we resort to {\it resampling}. Specifically, we consider either a {\it bootstrap} or a
{\it random permutation} sample of the original data and {\it then} apply the previous methodology.
 The resampled data behaves, asymptotically, as an independent sequence with
{\em unit} extremal index,
 that yields a second set of {\em estimating equations} of the tail index $\alpha$ and the parameter $\sigma$.
 By combining the resulting two estimating equations, one based on the original data and another based
 on the resampled data, we obtain a numerically stable estimate of $\theta$.

The resulting estimators for $\theta$ are shown to be {\em consistent} and {\em asymptotically normal}
for $m-$dependent sequences, while at the same time exhibiting good mean squared error properties in finite samples.
An additional advantage of resampling is that it provides a supplementary way of calculating confidence
intervals for $\theta$.  Resampling yields also new statistical plots, which provide further diagnostic tools for
quantifying the clustering of extremes at various magnitudes.  Simulation studies show that the
proposed estimator is a competitive alternative to existing ones. Further, it provides new insights at the important
parameter $\theta$ from the perspective of resampling, it provides new graphical tools, that can be successfully used
to analyze small as well as large data sets in practice.

The remainder of the paper is organized as follows: Section \ref{s:proposed-estimator} describes the proposed estimator.
Its asymptotic properties are established in Section \ref{s:theoretical-properties}.  Several methodological and algorithmic
issues are discussed in Section \ref{s:implementation}, while Section \ref{s:performance} focuses on
the evaluation of the estimator through an extensive simulation study.  Three important data sets of daily Crude Oil prices,
the daily returns of the S\&P 500 stock index, and the high--frequency traded volumes of the Intel stock are examined in Section
\ref{s:data-analysis}. The proofs and some auxiliary results are given in the Appendix.


\section{The max--spectrum based estimator of $\theta$ }
\label{s:proposed-estimator}

Let $X=\{X_k\}_{k \in \mathbb{Z}}$ be a positive ergodic strictly stationary sequence with
\emph{heavy tailed} marginals and {\em positive} extremal index $\theta>0$. Specifically, assume that
$ \P\{ X_k > x \} = 1-F(x) \sim  c_X x^{-\alpha}, \ \mbox{ as }x\to\infty$
for some $\alpha>0$ and $c_X>0$, where $a_n \sim b_n$ means $a_n / b_n \rightarrow 1$, as $n \rightarrow \infty.$
The parameter $\alpha$ corresponds to the \emph{tail index} of the distribution. Given a sample path $ X_1, \dots, X_n$,  we define the dyadic block maxima as follows:
\begin{equation} \label{equ:def_of_BM}
D(j,k) := \max_{1\le i \le 2^j} X_{{2^j}(k-1)+i} \equiv \bigvee_{i=1}^{2^j} X_{{2^j}(k-1)+i}\, ,
\end{equation}
where $j = 1,\ldots, \lfloor \log_2 n \rfloor $,  $k=1,\ldots,\lfloor  n/2^{j} \rfloor $, and where $\lfloor\cdot\rfloor$ denotes
the integer part function.
For heavy--tailed $X_k$'s, relation (\ref{e:M_n-iid-M_n}) holds with $H(x) = \exp\{ - c_X x^{-\alpha}\},\ x>0$
and normalization constants $c_n := n^{1/\alpha}$ and $d_n:= 0$. Therefore,
\begin{equation} \label{equ:convergence_of_D(j,k)_to_Frechet}
2^{-j/\alpha}D(j,k) \stackrel {D}{\longrightarrow}  \theta^{1/\alpha} \sigma Z^{1/\alpha},\ \ \mbox{ as }j\to\infty,  \ \ \mbox{ for fixed } k.
\end{equation}
where $Z$ is a standard $1-$Fr\'echet random variable, i.e.\ $\P\{Z\le z\}=\exp(-z^{-1})$, $z > 0$,
and where $\sigma:= c_X^{1/\alpha}$ is the asymptotic scale coefficient of the $X_k$'s.
Due to the nature of the Fr\'echet extreme value distribution, the extremal index parameter $\theta$
appears in the scale coefficient of the limit distribution of the dependent maxima. This feature
will play an important role in the estimation of $\theta$ discussed below.

Next, introduce the statistics
\begin{equation} \label{equ:Yj_definition}
 Y_j := \frac{1}{n_j} \sum_{k=1}^{n_j} \log_2(D(j,k)).
\end{equation}
where $n_j= \lfloor n/2^{j} \rfloor$. The statistics $Y_j,\ j=1\ldots, \lfloor \log_2(n) \rfloor$ will be referred to
as the \emph{max--spectrum} of the data, and the $j$'s as \emph{scales}.
By the assumed ergodicity and provided that moments exist, for a fixed $j$, we get
\begin{equation}
 Y_j \stackrel {a.s.}{\longrightarrow} \E Y_j  = j/\alpha + \E \log_2(2^{-j/\alpha}D(j,k)),\ \ \ \ \mbox{ as } n\to\infty.
\end{equation}
Assuming uniform integrability, relation (\ref{equ:convergence_of_D(j,k)_to_Frechet}), on the other hand, implies that
\begin{equation} \label{equ:Yj_regression_ss}
 \E Y_j \simeq  j/\alpha + \log_2(\sigma) + \E \log_2(Z) /\alpha  + \log_2(\theta)/\alpha,\ \ \ \ \mbox{ as } j \to \infty,
\end{equation}
where $a_n \simeq b_n$ means $a_n - b_n \rightarrow 0$, as $n \rightarrow \infty.$
This indicates the existence of a \emph{linear} relationship between the statistics $Y_j$ and $j$
up to an error term, which becomes negligible as $n_j$ and $j$ grow. The slope of a linear fit of $Y_j$ versus $j$
yields an estimator of $1/\alpha$ and thus $\alpha$. Although our goal is to estimate $\theta$,
the estimation of the tail index $\alpha$ is an intermediate step and an integral part of our analysis.

Observe that on the other hand for iid data, we have $\theta = 1$ and thus (\ref{equ:Yj_regression_ss}) becomes:
\begin{equation} \label{equ:Yj_regression_iid}
 \E Y_j^{iid}  \simeq  j/\alpha + \log_2(\sigma) + \E \log_2(Z)/\alpha ,
\end{equation}
where $\{Y_j^{iid}\}$ is the max--spectrum of an iid data set with the same distribution as the $X_k$'s.
Relations (\ref{equ:Yj_regression_ss}) and (\ref{equ:Yj_regression_iid}) suggest a method to obtain an estimate of $\theta$.
Namely, \emph{resample} the data, for example, by randomly drawing (with or without replacement) a sample
$X_1^*,\ldots,X_k^*$ of size $k=k(n)$ from the set $\{X_1,\ldots,X_n\}$.  Intuitively, this destroys the dependence
structure of the data, resulting in an approximately independent sample with the same marginal distribution
as the original stationary sequence.

Let $Y_j^*$ be as in (\ref{equ:Yj_definition}) where now the $D(j,k)$'s are based on the resampled data
$X_1^*,\ldots, X_k^*$. Since for an iid sequence we have $\theta=1$, we expect the resampled sequence to have
 $\theta \approx 1$, whereas $\alpha$ and $\sigma$ will remain unchanged. Thus, relation (\ref{equ:Yj_regression_ss})
becomes
\begin{equation} \label{equ:Yj_regression_shuffled}
 \E[Y_j^{*}] \simeq  j/\alpha + \log_2(\sigma) + \E[\log_2(Z)]/\alpha,
\end{equation}
where the term $\log_2(\theta)/\alpha$ is no longer present since $\log_2(\theta \approx 1) \approx 0$.

Thus, in view of (\ref{equ:Yj_regression_ss}) and (\ref{equ:Yj_regression_iid}), we have
$$
 Y_j^{*} \approx j/\alpha + \log_2(\sigma) + \E \log_2(Z) /\alpha,\ \ \ \mbox{ and }\ \ \
 Y_j \approx j/\alpha + \log_2(\sigma) + \E \log_2(Z) /\alpha + \log_2(\theta)/\alpha.
$$
Taking the difference between the last two estimating equations, replacing $\alpha$ by its estimate $\hat{\alpha}$ based
on (\ref{equ:Yj_regression_ss}), and solving for $\theta$ we obtain the following estimator for the extremal index:
\begin{equation} \label{equ:main_definition_of_theta}
\hat{\theta}(j) = 2^{-\hat{\alpha}(j)(Y_j^{*} -  Y_j)}.
\end{equation}

\begin{figure}[th!]
\begin{center}
\includegraphics[width=2.5in]{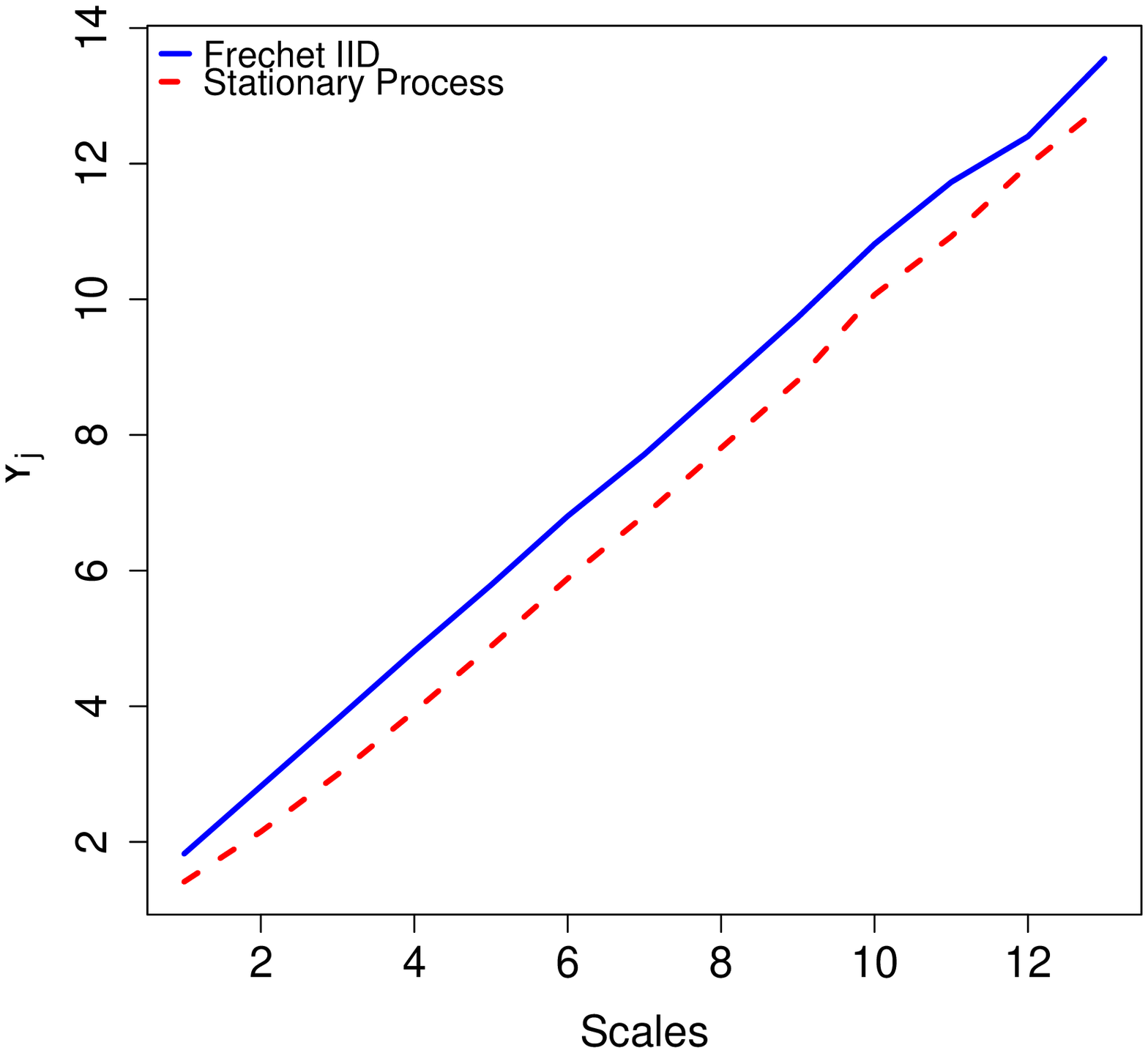}\includegraphics[width=2.5in]{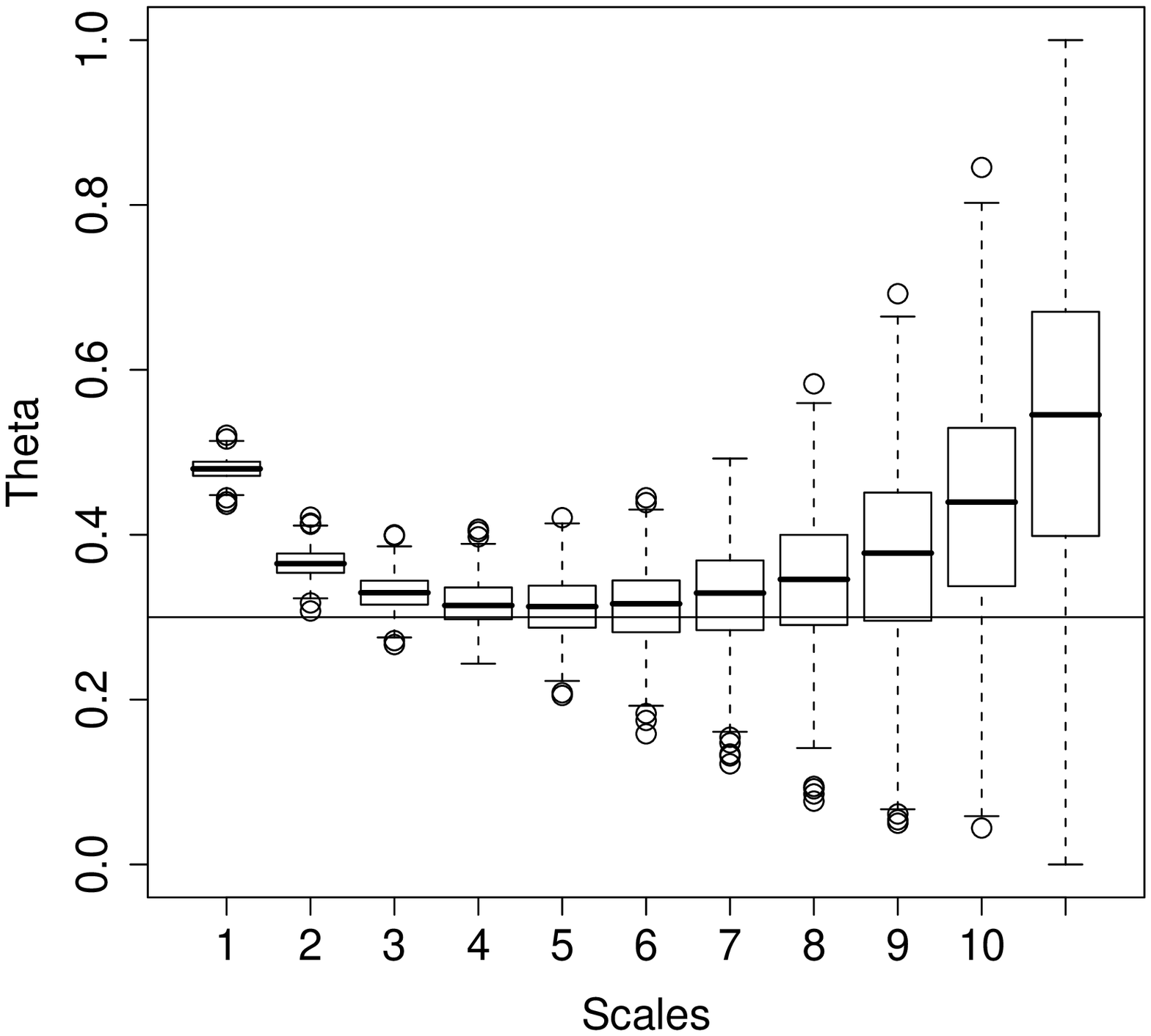} 
{\caption{\label{StationaryExample} {\it Left panel:} The max--spectrum
 of $X_n = \max\{\frac{2}{3}X_{n-1}, \frac{1}{3}Z_n\}$, $\theta=1/3$, with  $Z_i$'s iid standard $1-$Fr\'echet
 ({\it solid line}) and the max--spectrum of iid copies of the $X_k$'s ({\it broken line}). The two spectra are essentially linear with
 equal slopes. {\it Right panel:} boxplots of the $\hat{\theta}(j)$'s obtained from different resampled versions of a single path
of the process.  The circles indicate outliers located more than 1.5 fourth--spreads away from the sample median and the horizontal line is the
theoretical value of $\theta=1/3$.}}
\end{center}
\end{figure}

Observe that for a single data set, one can obtain a large set of estimates $\hat \theta(j)$, based on different
resampled versions of the data. Thus, resampling allows us to gauge the variability of the
estimates as well as the range of scales $j$ where the asymptotics in  (\ref{equ:Yj_regression_ss})
and (\ref{equ:Yj_regression_iid}) become applicable.

Figure \ref{StationaryExample} illustrates the main principle behind the proposed estimator.
The left panel shows the combined max--spectra of a dependent sequence and an iid sample. The two max--spectra are
parallel with equal slopes $\approx 1/\alpha$, since the marginal distributions behind the two spectra are the same. The
difference is in the \emph{intercept} and this is where the value of $\theta$ is derived from.  The right panel shows
boxplots of $\hat \theta(j)$ estimates obtained from $200$ independent resampled versions of a single path of the
process on the left.  Observe that the medians of the $\hat \theta(j)$'s closely follow the true value $\theta = 1/3$
over a range of scales (for more details, see Section \ref{s:implementation} below).

\noi{\em Remarks:} \\
(1) The statistics $Y_j$'s in (\ref{equ:Yj_definition}) are not only dependent in $j$, but more importantly, they have
 different variances in $j$ since they involve averages of $n_j \approx n/2^j$ terms. Thus, to reduce the variance in
 the regression estimators of $\alpha$, it is essential to use a weighted or generalized least squares method
 (see e.g.\ \cite{stoev1}, for more details). \\
(2) The proposed resampling procedure avoids the problem of estimating the scale parameter $\sigma = c_X^{1/\alpha}$,
 however, an estimate of $\alpha$ is still needed. The algorithmic implementation of the estimators $\hat \theta(j)$
and other important practical issues are discussed below.
The appropriate resampling sample size $k(n)$, from the perspective of asymptotics, is $o(\sqrt{n})$
(see, Section \ref{s:theoretical-properties}). \\
(3)  The estimate $\hat\theta(j)$ depends on the scale $j$, as indicated. An {\it automatic procedure} for the choice
 of $j$ is presented in Section \ref{s:implementation}.


\section{Theoretical properties}
\label{s:theoretical-properties}


Let $X = \{X_k\}_{k \in \mathbb{Z}}$ be a strictly stationary time series with marginal heavy-tailed c.d.f.\ $F$
and let also $M_n =\max_{1\le i \le n} X_i \equiv  \bigvee_{i=1}^n X_i$. We then have
\begin{equation} \label{e:F_n}
 F_n(x) := \P\{  M_n \le  n^{1/\alpha} x\} = \exp\{ - c(n, x) x^{-\alpha}\},\ \ x\in \mathbb{R},
\end{equation}
for some function $c(n,x)>0$, $n\in \mathbb{N}$.
As in (\ref{equ:convergence_of_D(j,k)_to_Frechet}), if the time series $X$ has a positive \emph{extremal index}
$\theta \in (0,1]$, then
\begin{equation}\label{e:Mn-to-theta}
 n^{-1/\alpha} M_n \stackrel{D}{\longrightarrow} (\theta c_X)^{1/\alpha} Z^{1/\alpha},\ \ \mbox{ as }n\to\infty,
\end{equation}
where $Z$  is a standard $1-$Fr\'echet variable: $\P\{Z\le x\} = e^{-x^{-1}},\ x>0$.

Our asymptotic results rely on the moment behavior of $f(M_n/n^{1/\alpha})$, for certain deterministic
functions $f$ and involve some additional technical conditions, outlined below (for more details,
see the Appendix).

\noindent\textbf{Condition 1.}  \emph{There exists $\beta>0$ and $R \in \bbR$, such that
\begin{equation} \label{e:C1}
|c(n,x) - \theta c_X| \le c_1(x) n^{-\beta},\ \ \mbox{ for all } x >0,\ \ \ \ \mbox{ and }\ \ \ c_1(x) = {\cal O}(x^{-R}),\ x\downarrow 0,
\end{equation}
where $\theta\in (0,1]$.}

\noindent\textbf{Condition 2.} \emph{$F_n(0) = 0$ and for all $x>0$,
\begin{equation} \label{e:C2}
 c(n,x) \ge c_2 \min\{1,x^\gamma\},\ \ \mbox{ for some }\gamma\in (0,\alpha),
\end{equation}
for all sufficiently large $n \in \mathbb{N}$, where $c_2>0$ does not depend on $n$.}

\noi {\em Remarks: } \\
 (1) The conditions (\ref{e:C1}) and (\ref{e:C2}) are not very stringent.  For example, let
\begin{equation}\label{e:simple-process}
X_k = \max\{ Z_k, Z_{k-1},\ldots,Z_{k-m+1}\},\ \ k\in \mathbb{Z},
\end{equation}
where the $Z_k$'s are independent, standard $\alpha-$Fr\'echet.  We then have
\begin{eqnarray*}
\P\{ M_n\le n^{1/\alpha} x\} &= & \P\{ Z_{-m+1}\le n^{1/\alpha}x,\cdots, Z_n \le n^{1/\alpha}x\}
 = \exp\{ - c(n,x) x^{-\alpha}\},
\end{eqnarray*}
where the function $c(n,x) = (n+m-1)/n = 1 + {\cal O}(1/n)$
does not depend on $x$ and $\beta = 1$, in this simple case.
Conditions 1 and 2 above hold for a more general class of moving maxima processes (see \cite{hamidieh:stoev:michailidis:2007}).  \\
(2) Condition 1 and relation (\ref{e:F_n}) imply (\ref{e:Mn-to-theta}), that is, the extremal index
 of the time series $X$ is precisely equal to $\theta$ in (\ref{e:C1}). Thus, (\ref{e:C1}) quantifies
 further the {\it rate} of the convergence in (\ref{e:Mn-to-theta}).

\noi{\bf Description of the asymptotic regime:} To obtain the consistency of statistics based on the max--spectrum $Y = \{Y_j\}$,
we focus on the range of scales $[j(n), \ell +j(n)],$ where $\ell \in \bbN$ is
fixed and where $j(n) \to \infty,$ as  $n\to\infty$.  We then define
\begin{equation}\label{e:alpha-j}
 \hat \alpha (j) := {\Big(} \sum_{ i = 0}^{\ell} w_i Y_{i+j(n)} {\Big)}^{-1},
\end{equation}
where the weights $w_i$'s are fixed and such that $\sum_{i=0}^{\ell} w_i = 0$ and $\sum_{i=0}^{\ell} i w_i =1.$
The weights $w_i$'s can be obtained, for example, either from GLS or WLS regression of
$Y_{i+j(n)}$ versus $i,$ for $0 \le i \le \ell$ (see \cite{stoev1}, for more details).

The estimator $\hat \theta$ in (\ref{equ:main_definition_of_theta}) involves both the max--spectrum $Y$
of the {\it dependent} data and the max--spectrum $Y^*$ of the {\it resampled} data. Observe that
\begin{equation}\label{e:theta-j}\label{e:C-j}
 \hat \theta (j) = 2^{ - \hat \alpha(j) (C^*(j) -  C(j))},\ \ \mbox{ where }\ \
  C^*(j) :=  Y^*_j - j/\alpha\ \ \ \mbox{ and  }\ \  \ \ \ C(j) := Y_j - j/\alpha,
\end{equation}
since trivially $Y^*_j - Y_j = C^*(j) - C(j)$. We will establish the asymptotic normality of $\hat \theta(j)$ in three steps:

{\it (Step 1.)} We first establish rates of convergence for the quantities $\hat \alpha(j)$
 and $C(j)$, which are based on the max--spectrum $\{Y_j\}$.

{\it (Step 2.)} We then show that the $C^*(j)$'s are asymptotically normal (under certain conditions) in two
resampling schemes: {\it bootstrap} and {\it random permutations}.

{\it (Step 3.)} We finally combine the results from {\it Steps 1.} and {\it 2.} above to establish the
asymptotic normality of $\hat \theta (j)$.

\noi{\bf Main results:} We establish next the asymptotic normality of  $\hat \theta(j)$ defined in (\ref{e:theta-j}),
by following the three steps outlined above.

\noi
{\bf Step 1:} The following result provides rates of convergence for $\hat\alpha(j)$ and $C(j)$.

\begin{prop} \label{p:alpha-C} Let $X_1,\ldots, X_n$ be a sample from an $m-$dependent, strictly stationary time
series $X = \{X_k\}_{k\in\bbZ}$, which satisfies Conditions 1 and 2 above.

Then, for $\hat\alpha(j)$ and $C(j)$ in (\ref{e:alpha-j}) and (\ref{e:C-j}), we have, as $n\to\infty$
\begin{equation}\label{e:p:alpha} \label{e:p:C}
 \hat \alpha(j) = \alpha + {\cal O}_P(\frac{1}{2^{j(n)\min\{1,\beta\}}}) + {\cal O}_P(\frac{2^{j(n)/2}}{n^{1/2}}),
\ \ \mbox{ and } \ \
 C(j) =  C + {\cal O}_P(\frac{1}{2^{j(n)\min\{1,\beta\}}}) + {\cal O}_P(\frac{2^{j(n)/2}}{n^{1/2}}),
\end{equation}
with
$
 C = \log_2(\theta)/\alpha + \log_2(c_X)/\alpha + \E \log_2(Z)/\alpha,
$
where $Z$ is a standard $1-$Fr\'echet variable.
\end{prop}

\noi The proof of this result is given in the Appendix.  Observe that Proposition \ref{p:alpha-C} is valid
for an arbitrary stationary $m-$dependent time series which satisfies (\ref{e:C1}) and (\ref{e:C2}). It is valid, in particular,
for the simple process $\{X_k\}_{k\in\bbZ}$ in (\ref{e:simple-process}) and more generally for the moving maxima processes in (\ref{equ:def_of_Xk}).

\noi{\bf Step 2:} We now employ resampling to obtain an approximately independent data
sample $X_1^*,\ldots,X_k^*$. Here, we consider two resampling schemes, the first based on {\it bootstrap} and
the second on {\it permutations}. We then establish asymptotic normality results for the max--spectrum in both schemes.
The sample $X_1^* := X_{i_1},\ X_2^*:= X_{i_2},\ldots, X_k^* := X_{i_k}$ is
a {\it bootstrap} sample from the data $X_1,\ldots, X_n$ if the indices $i_1,\ldots,i_k$ are drawn randomly and
with replacement from the set $\{1,\ldots,n\}$.  When these indices are drawn without replacement and $k\le n$,
we obtain a {\it permutation} sample. We need the following:

\begin{lemma}\label{l:boot-permute} Let $i_1,\ldots, i_k$ be a collection of randomly drawn indices
either with replacement or without replacement from the set $\{1,\ldots,n\}$.
For any fixed $m\in\bbN$, we have
$$
 \P\{ \min_{1\le j'< j''\le k} |i_{j'} - i_{j''}| \ge m\} \ge 1 - m k^2/(n-k).
$$
\end{lemma}

\noi The proof is given in the Appendix. This result implies that for
$k(n) = o(\sqrt{n}),\ n\to\infty$, the indices $\{i_j,\ 1\le j\le k\}$ are spaced by at
least $m-$lags away from each other, with probability asymptotically equal to $1$, as $n\to\infty$.
Therefore, if the data $X_1,\ldots,X_n$ come from an $m-$dependent time series,
for the purposes of asymptotics in distribution, both the {\it bootstrap} and the {\it permutation} samples
of size $k = o(\sqrt{n})$ become essentially independent, with high probability, as $n\to\infty$.
This fact and Proposition 4.2 in \cite{stoev1}, readily imply the following result.

\begin{theorem} \label{t:C-B} Let $X=\{X_i\}_{i\in\bbZ}$ be a strictly stationary $m-$dependent time series,
which satisfies Conditions 1 and 2 above. Let $X_1^*,\ldots, X_k^*$ be either a bootstrap or a permutation
sample from $X_1,\ldots, X_n$, where $k(n)\to\infty$ is such that $k(n) = o(n^{1/2}),$ as $n\to\infty$,
and let $Y^*$  be its corresponding max--spectrum.

Let $j(k)\to\infty,\ n\to\infty$, be such that $k/2^{j(k)(1+2\beta)} + j(k)^2 2^{j(k)}/k \longrightarrow 0,$
as $k\to\infty$.

\noindent Then, for $C^*(j)$ in (\ref{e:C-j}), we have
\begin{equation}\label{e:C-B-limit}
 \sqrt{k_{j}} (C^*(j) - C^*) \stackrel{D}{\longrightarrow} {\cal N} (0, \sigma_{C^*}^2),\ \ \ \mbox{ as }n\to\infty,
\end{equation}
where  $k_j = k(n)/2^{j(n)}$. Here $C^* := \log_2(c_X)/\alpha + \E \log_2(Z)/\alpha$, and
$\sigma_{C^*}^2 =  \alpha^{-2}{\rm Var}( \log_2  Z  ),$ where $Z$ is a standard $1-$Fr\'echet variable.
\end{theorem}

\noi
The proof is given in the Appendix.

\noindent{\bf Step 3:} The following Theorem is the main result of this Section.
 It combines the results of Proposition \ref{p:alpha-C} and Theorem \ref{t:C-B} to establish
the asymptotic normality of $\hat \theta (j)$.

\begin{theorem}\label{t:main} Assume the conditions of Theorem \ref{t:C-B} and let $\hat \alpha(j)$ be as
in (\ref{e:alpha-j}), where $Y$ is the max--spectrum of the data $X_1,\ldots,X_n$.
Let also $C(j)$ and $C^*(j)$ be as in (\ref{e:C-j}), where $Y^*$ is the max--spectrum of either
a bootstrap or a permutation sample $X_1^*,\ldots, X_k^*$ of the data.

Let $k(n) = o(\sqrt{n}),\ n\to\infty$ and $j(k)\to\infty,\ k\to\infty$, be such that
\begin{equation}\label{e:k(n)-main}
 k /2^{j(k) (1+2\min\{1,\beta\})} + j(k)^2 2^{j(k)}/k \longrightarrow 0,\ \ \ \mbox{ as }k\to\infty,
\end{equation}

\noi Then, for $\hat\theta(j)$ in (\ref{e:theta-j}), we have
$$
\sqrt{k_{j} } (\hat \theta(j) - \theta) \stackrel{D}{\longrightarrow} {\cal N}(0,\theta^2 \pi^2/6),
\ \ \ \mbox{ as }n\to\infty,
$$
where $k_j = k(n)/2^{j(n)}$.
\end{theorem}

\noi The proof of this result is given in the Appendix. A few important remarks follow.

\noi{\em Remarks}\\
 (1) Theorem \ref{t:main} applies, for example, to the class of moving maxima processes
 in (\ref{equ:def_of_Xk}), under mild assumptions on the innovations
 $Z_k$'s (see Conditions $1'$ \& $2'$ below). It holds, for example, for
 Pareto, mixtures of Pareto or Fr\'echet innovations. \\
 (2) Let $\delta \in (0, 2\min\{1,\beta\})$ be arbitrary and suppose that $k/2^{j(k)(1+2\min\{1,\beta\})} \sim k^{-\delta},\ k\to\infty$.
 We then have $2^{j(k)} \sim k^{(1+\delta)/(1+2\min\{1,\beta\})},\ k\to\infty$ which, since $\delta < 2\min\{1,\beta\}$, implies
 that relation (\ref{e:k(n)-main}) holds.
 This yields the rate $k_j \sim k^{(2 \min\{1,\beta\} + \delta)/(1+2\min\{1,\beta\})}$ in Theorem \ref{t:main}. Since
 $k=o(\sqrt{n})$ and since $\delta>0$ can be taken arbitrarily small, we can achieve rates up to $n^{\frac{\min\{1,\beta\}}{(1+2\min\{1,\beta\})}}$.
 For example, if $\beta>1/2$ the rate of $n^{1/4}$ is possible while the best possible rate is $o(n^{1/3})$.


\section{Implementation issues}
 \label{s:implementation}

We present next an algorithmic implementation for the proposed estimator of $\theta$ and discuss its main
features.  We then propose a second algorithm for the automatic selection of scales.

In Theorem \ref{t:main}, we only consider resampled sets from the data of size $k(n) = o(\sqrt{n})$. In practice,
we found that the estimators of $\theta$ continue to work well even if one considers random permutations
of the entire data sample of size $k(n) = n$. Using bootstrap instead of permutation samples, results in estimates
$\hat \theta (j)$ with larger variances and bias (for large $j$'s), especially for small sample sizes. Thus,
in the sequel, we focus on permutation based resampling and utilize the entire data set.

\noi {\bf Algorithm 1:} {\it (estimation of $\theta$)}

\begin{enumerate}

 \item Compute the $Y_j$'s and the $\hat{\alpha}(j)$'s as in (\ref{equ:Yj_definition}) and (\ref{e:alpha-j}) based on the original data.

 \item Randomly permute (i.e.\ shuffle) the data, $N_{in}$ times and collect the $N_{in}$ statistics $Y_j^{*}$.

 \item Find the $N_{in}$ differences of $Y_j^{*} - Y_j$ and compute the sample mean for the positive differences only:
   $\Delta(j) = {\rm mean}\{Y_j^{*} - Y_j\}_{+}.$

 \item Obtain the estimates of $\theta$ for each scale $j$: $\hat{\theta}(j) = \max\{2^{-\hat{\alpha}(j) \Delta(j) },1\}.$

 \item Repeat steps 2, 3, and 4, $N_{out}$ number of times and collect the $\hat{\theta}(j)$ values.

 \item Produce a sequence of $\hat{\theta}(j)$ boxplots from the $N_{out}$ available values, per each scale $j$.

 \item Visually inspect the boxplots of $\hat{\theta}(j)$ and select a range of scales where the \emph{medians} of the
    boxplots stabilize. Estimate $\theta$ by using the median values from this range of scales.
\end{enumerate}

\medskip
\noi In the following remarks we explain and justify the steps in the above algorithm.

\medskip
\noi {\bf Discussion of Algorithm 1:}

{\it Step 1:} The estimate $\hat{\alpha}(j)$ is based on the range of scales $j, \ldots, j+\ell$, where
$j+\ell = \lfloor \log_2 (n) \rfloor -1$ is chosen to be the second largest available scale in the data.
In practice, we discard the highest scale since it involves an average of at most two block--maxima.
We recommend using either generalized least squares with the asymptotic covariance matrix for the max--spectrum
given in \cite{stoev1} or weighted least squares which account for the fact that ${\rm Var}(Y_j) \propto 1/n_j \propto 2^j$.
Both approaches are comparable and considerably better than ordinary least squares regression, which should not be used.

{\it Steps 2 \& 3:} We introduce an \emph{inner loop} with $N_{in}$ iterations to reduce the variability of $Y_j^{*} - Y_j$.
This considerably improves the variance of the $\theta$ estimates.
On step 3, we average only the positive differences $Y_j^{*} - Y_j$ since by relations (\ref{equ:Yj_regression_ss})
and (\ref{equ:Yj_regression_shuffled}), we have $\E Y_j^{*} \geq \E Y_j$.
Our experiments indicate that replacing the ``mean'' by ``median'' in step 3 yields similar results.

{\it Step 4:} As in \cite{ferro-segers}, we take the minimum of the calculated estimate and 1 to ensure that $\hat \theta(j) \in [0,1]$.

{\it Step 5:} This step yields a sample of $N_{out}$ estimates of $\theta$ for each scale $j$.
The practical choice of the parameters $N_{out}$ and $N_{in}$ is discussed in Section \ref{s:performance}.

\begin{figure}[t!]
\begin{center}
\includegraphics[width=2.5in]{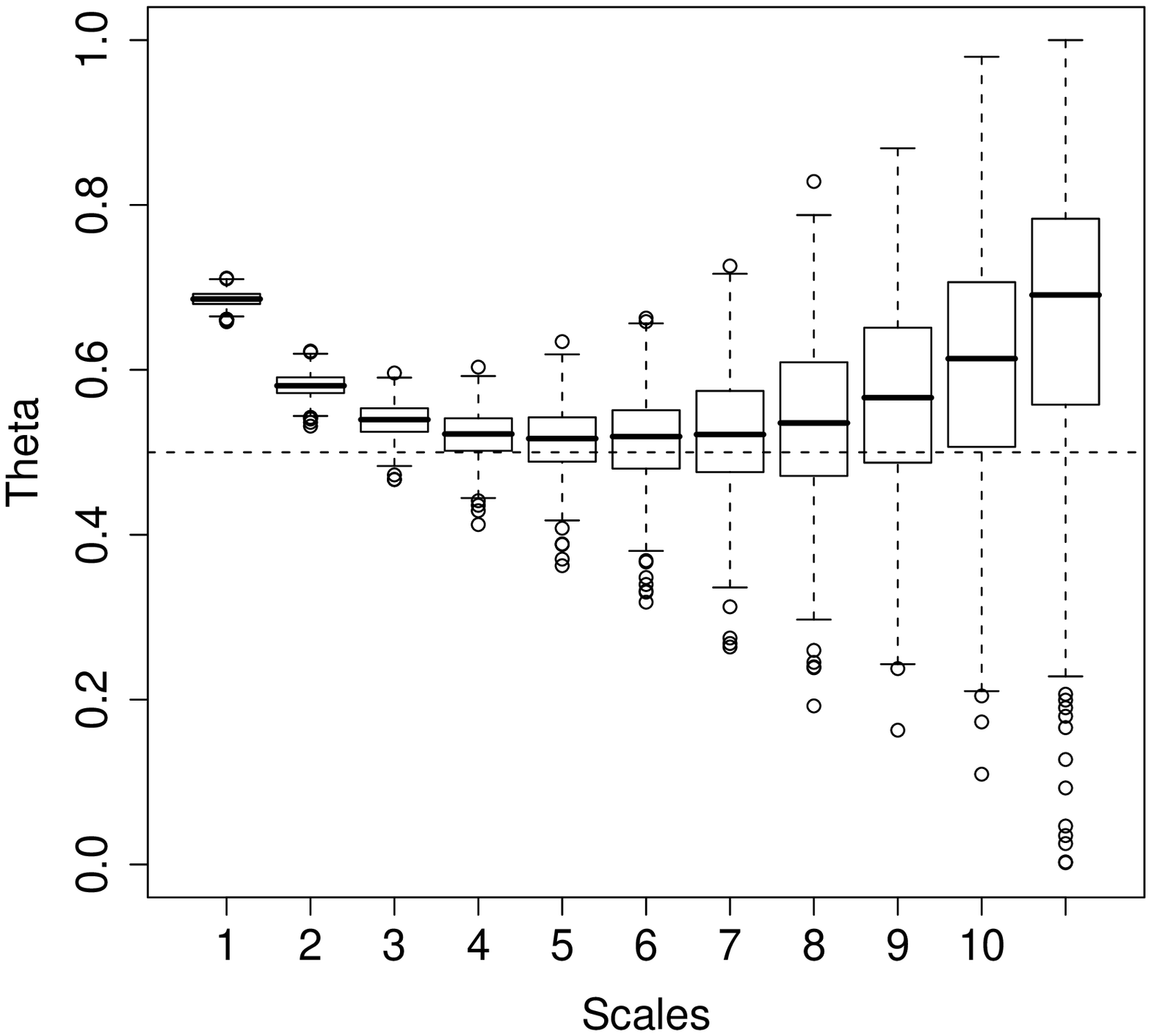}\includegraphics[width=2.5in]{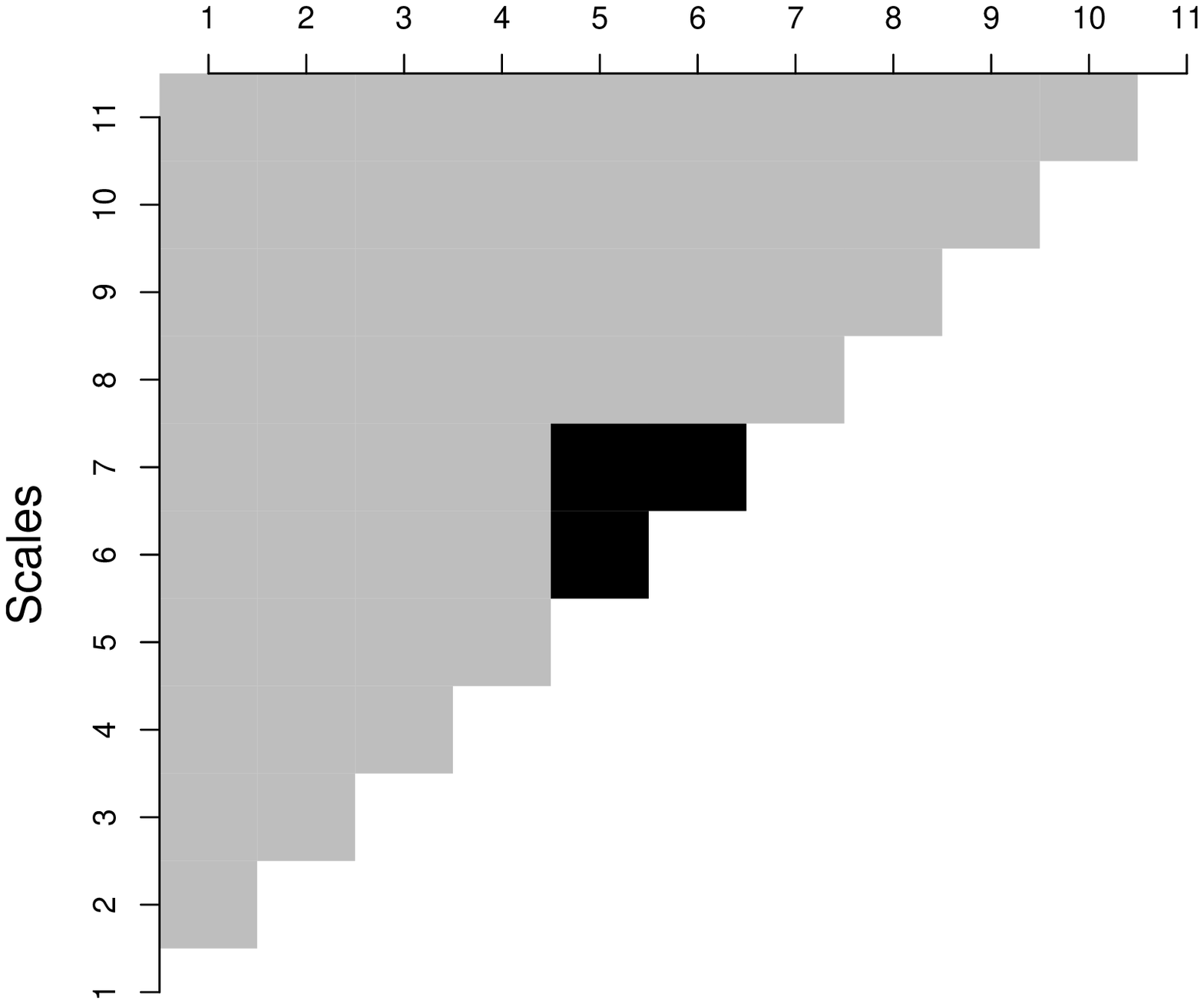}
{\caption{\label{fig:heatmap} Estimation of $\theta$ for the process
 $X_n = \max\{\frac{1}{2}X_{n-1}, \frac{1}{2}Z_n\}$, $\theta=1/2$, $Z_i$ iid standard $1-$Fr\'echet, with
sample size of n=$2^{13}$, $N_{out}=500$ and $N_{in}=25$.  {\it Left panel:} Boxplots of $\hat{\theta}(j)$'s with the last two scales omitted.
{\it Right panel:} A 'heat map' visualizing the Kruskal--Wallis test for the automatic selection of scales -- black corresponds to
 $p-$values greater than $0.05$. }}
\end{center}
\end{figure}

{\it Step 6:} In practice, the estimation of $\theta$ requires selecting the range of scales, where
the best bias/variance trade--off is achieved.  Estimating $\theta$ over the larger scales $j$
(larger block sizes) involves lower bias, but leads to larger variance as the number of block--maxima
 is reduced.  At lower scales $j$ (smaller block sizes) the bias grows but the variance is reduced (see Figure \ref{fig:heatmap}).
 In general, reliable estimates of $\theta$ can be obtained from the middle range of scales.
 The choice of the scales $j$ is addressed in the sequel.

\medskip
Figure \ref{fig:heatmap} (left panel) illustrates the above algorithm over a simulated process with known
extremal index $\theta=1/2$.  A stable range of scales 4 to 7 can be observed.
In practice, we recommend taking the median of the sample of the pooled $N_{out}$ estimates $\hat \theta(j)$
from each one of the scales $j$ in the stable range. In this case we obtained a point estimate of $0.52$.
One can also obtain an empirical 95\% confidence interval, based on 0.025--th and 0.975--th empirical quantiles
of the pooled $\hat \theta(j)$ values to obtain $(0.40, 0.62)$ (see also relations (\ref{e:CI_based_on_normality}) and
(\ref{e:CI_based_on_quantiles}) below).

\medskip
The selection of the stable range of scales $j$ in Step 6 of Algorithm 1 is subjective.
We propose next an automated procedure for selecting the range of scales, based on the Kruskal--Wallis test.

\medskip
\noi{\bf Algorithm 2:} {\it (automatic selection of scales)}

\begin{enumerate}
 \item For every given range $j_1\le j \le j_2,\ j_1<j_2$ of possible consecutive scales in the data, perform
 a Kruskal--Wallis test for equality of the medians, based on the samples of $N_{out}$ values of $\hat\theta(j)$.

 \item Consider the array of $p-$values: $p(j_1,j_2)$ resulting from the tests in Step 1. Declare the medians over
  the range $[j_1,j_2]$ 'statistically different' if $p$ is less than a prescribed {\it significance threshold}.

 \item Produce a pooled estimate of $\theta$ based on the longest scale range where the medians are 'statistically equal'.

 \item If there are ties in Step 3, pick the range starting at the lowest scale. If all medians are 'statistically
  different', pick the middle scale and follow up by a visual inspection of the results.

\end{enumerate}

\medskip
\noi The proposed automatic scale selection procedure is evaluated in Section \ref{dd}.
One possible method to visualize the results of this analysis is to construct a  'heat map' of the p-values
for the Kruskal--Wallis tests -- see Figure \ref{fig:heatmap} (right panel).
The axes correspond to scales $j_1$ and $j_2$; the regions in black  indicate ranges of scales $[j_1,j_2]$
with $p-$values greater than $0.05$. This heat map shows that the medians over the scale range $[j_1,j_2]=[5,7]$
are 'statistically equal' at a level of $5\%$.  A point estimate based on the pooled values
from scales 5 to 7 is 0.52 with an empirical 95\% confidence interval of (0.39, 0.63).


\section{Performance evaluation} \label{dd}
 \label{s:performance}

We present next the results of a simulation study and comment on the performance of the
max-spectrum, the {\it Ferro-Segers} (\cite{ferro-segers}) and the {\it runs} (\cite{obrien:1987}) estimators for
the extremal index. We briefly summarize these two competing estimators next:

The first estimator is based on the characterization of the
extremal index given by \cite{obrien:1987}. In this characterization, $\theta$ is expressed as the limiting probability
that an exceedance is followed by a run of observations below a high threshold $u_n$:
$$\theta = \lim_{n \rightarrow \infty} P\{ \bigvee_{j=2}^{r_n}X_j
\leq u_n | X_1 > u_n \},$$
where $r_n=o(n)$ is the \emph{length of runs} of values of the process falling below the threshold given that an exceedance has occurred.
This characterization motivates the definition of the \emph{runs} estimator for a fixed high threshold $u$ and a specified runs length
$r$:
\begin{equation} \label{equ:runs_definition}
\hat{\theta}_{runs} = \frac{\sum_{j=1}^{n-r}\mathbf{I}(X_j \geq u
\geq \bigvee_{i=j+1}^{j+r}X_i)}{\sum_{j=1}^{n-r}\mathbf{I}(X_j > u)}.
\end{equation}
The runs estimator is asymptotically normal and consistent.
See \cite{weissman:novak:1998} and references therein for additional information.

The second estimator is due to \cite{ferro-segers}.  An interesting aspect of this estimator is
that it does not require an auxiliary parameter (run
length in the case of the runs estimator).  However, one still has to choose the
threshold.  Using a point process approach, \cite{ferro-segers} show that the
inter-exceedance times - time differences between successive values above a threshold - of the extreme values normalized by
$\bar{F}(u_n)$ converge in distribution to a random variable
$T_{\theta}$ with a mass of $1-\theta$ at $t = 0$ and an exponential
distribution with rate equal to $\theta$ on $t>0$.
Using a moment estimator, they
first obtain:
$$\hat{\theta}_1 =
\frac{2(\sum_{i=1}^{N-1}T_i)^2}{(N-1)(\sum_{i=1}^{N-1}T_i^2)},$$
where $\{T_i\}$ are the inter-exceedance times and $N$ is the number
of exceedances of a fixed high threshold $u$. A bias corrected version gives,
$$\hat{\theta}_2 =
\frac{2(\sum_{i=1}^{N-1}(T_i-1))^2}{(N-1)(\sum_{i=1}^{N-1}(T_i-1)(T_i-2))}.$$
To obtain the final form of the
estimator, a further adjustment is made to ensure that the values of the
estimator lie between 0 and 1:
\begin{equation}
 \hat{\theta}_{F/S} = \left\{ \begin{array}{ll}
  1 \wedge \hat{\theta}_1 & $if $\max\{T_i: 1 \leq i < N - 1\} \leq 2, \\
  1 \wedge \hat{\theta}_2 & $if $\max\{T_i: 1 \leq i < N - 1\} > 2. \\
\end{array} \right.
\end{equation}
The Ferro-Segers estimator is consistent for $m$-dependent strictly stationary sequences.

Next, we discuss three types of processes, used in the simulation study, for which the
extremal index is given in closed form.

\noindent
$\bullet$ The \emph{max-autoregressive} (armax) process of order one is defined as:
$$X_n = \max\{bX_{n-1}, (1-b)Z_n\},\ \  \ \mbox{ where }\ \  0 \le b <1, $$
and where $\{Z_n\}_{n \in \bbZ}$ is an iid sequence of standard $\alpha-$Fr\'echet random variables.
For such processes $\theta = 1 - b^\alpha$ can  take any value in the interval $(0,1]$
(see e.g.\ \cite{beirlant:goegebeur:segers:teugels:2004} for additional information).

\noi$\bullet$ The \emph{linear process} $\{Y_n\}$, $n \in \bbZ$ is defined as:
$$Y_n = \sum_{j \in \bbZ} \psi_j Z_{n-j}, \ n \in \bbZ,\ \ \ \mbox{ where }\
 \sum_{j\in\bbZ} |\psi_j|^\delta < \infty,\ \ \mbox{ for some
}0<\delta<\min\{1,\alpha\}. $$
Here $\{Z_n\}_{n\in\bbZ}$ is an iid sequence of heavy--tailed innovations with exponent $\alpha>0$.
When the $Z_n$'s are symmetric, we have
$\theta = (\psi_{+}^{\alpha} + \psi_{-}^{\alpha}) \ / \|\psi\|_{\alpha}^{\alpha},$
where $\psi_{+}= \max_{j}(\psi_j \vee 0)$, $\psi_{-}= \max_{j}(-\psi_j \vee 0)$, and
$\|\psi\|_{\alpha}^{\alpha} = \sum_{j\in\bbZ} |\psi_j|^{\alpha}$
(see, e.g.\ Corollary 5.5.3 in \cite{embrecht:kluppelberg:mikosch:1997}).  We will use iid t-distributed innovations
$Z_n$'s where the degrees of freedom parameter is also equal to the tail index $\alpha$.

\noi$\bullet$ The {\it moving maxima process} $X = \{X_k\}_{k\in\bbZ}$ is defined as:
\begin{equation} \label{equ:def_of_Xk}
  X_k := \max_{1\le i\le m} a_i Z_{k-i+1}, \ \ \ k\in \bbZ,
\end{equation}
with some coefficients $a_i>0,\ i = 1,\ldots, m,$ and $m\ge 1$, where the $Z_k$'s are iid,
positive heavy--tailed random variables with tail exponent $\alpha$.  The extremal index $\theta$
of $X$ is: $\theta = \max_{1\le i \le m} a_{i}^\alpha/ \sum_{i=1}^m a_i^\alpha$.

\noi{\bf Simulation setup:} For brevity, we present selected results for the processes under consideration that
demonstrate best the behavior of the various estimators.

  $\circ$ $X_n = \max\{bX_{n-1}, (1-b)Z_n\}$, with $Z_i$ iid standard $1-$Fr\'echet.

  $\circ$ $Y_n = 0.50Z_n + 0.20Z_{n-1} + 0.10Z_{n-2}$, with $Z_i$ iid t-distributed with $\alpha$ degrees of freedom.

  $\circ$ $W_n = \max\{0.80Z_n, 0.20Z_{n-1}, 0.40Z_{n-2}\}$, with $Z_i$ iid Pareto with tail index $\alpha$.



\noi$\bullet$ {\it Parameters:} For the armax processes, we fix the tail index at $\alpha=1$ and vary the
coefficient $b$ to obtain a range of $\theta$ values.  The coefficients of the linear and moving maxima processes
are fixed (as indicated above), and the values of $\alpha$ for the $Z_k$'s are varied to obtain a range of $\theta$
values.  For all processes, other choices of the parameters produced analogous results.
For each type of process, 500 independent sample paths were generated
of length $2^{13} = 8192$ for the armax and moving max processes and $2^{14} = 16384$
for the linear processes.
For each generated sample path, the Ferro--Segers, the runs 1, 5, and 9 at each
selected threshold were computed. The proposed {\it max--spectrum} based estimator was computed using both GLS and WLS
and setting $N_{in}=25$. The threshold (Ferro-Segers and runs estimators) and the scale
(proposed estimator) parameters achieving the \emph{best} Root-Mean-Square-Error (RMSE) are reported
in Tables \ref{table:armax-table} -- \ref{table:moving-maxima}.

The results  demonstrate that the proposed {\it max--spectrum} estimator exhibits a good overall performance
in terms of RMSE and in many settings it outperforms the {\it Ferro-Segers} estimator. The GLS and WLS
variants produce similar results. The {\it runs} estimator performs exceptionally well for the armax
process, {\em if} the 'correct' run-length parameter is specified. However, it is quite sensitive to the
type of process and to the choice of the run-length parameter employed. The {\it max--spectrum} and {\it Ferro-Segers}
estimators are significantly more robust than the {\it runs} estimator to the choice of the model.

\begin{table}[h!]
\scriptsize
\def\baselinestretch{1.5}
\begin{center}
\begin{tabular}{|c|c|c|c|c|c|c|c|}
  \hline
$\theta$  &  $\alpha$ &   $GLS$   & $WLS$     & $F/S$     & $Runs-1$  & $Runs-5$  & $Runs-9$ \\ \hline \hline
0.10      &   1.00    &   0.0189  &   0.0197  &   0.0140  &   \textbf{0.0109}  &   0.0127  &   0.0137  \\ \hline
0.20      &   1.00    &   0.0226  &   0.0256  &   0.0206  &   \textbf{0.0164}  &   0.0218  &   0.0247  \\ \hline
0.30      &   1.00    &   0.0325  &   0.0291  &   0.0272  &   \textbf{0.0223}  &   0.0298  &   0.0343  \\ \hline
0.40      &   1.00    &   0.0334  &   0.0290  &   0.0306  &   \textbf{0.0272}  &   0.0381  &   0.0440  \\ \hline
0.50      &   1.00    &   0.0335  &   0.0308  &   0.0316  &   \textbf{0.0302}  &   0.0436  &   0.0520  \\ \hline
0.60      &   1.00    &   0.0350  &   \textbf{0.0310}  &   0.0326  &   0.0316  &   0.0485  &   0.0569  \\ \hline
0.70      &   1.00    &   0.0323  &   \textbf{0.0285}  &   0.0348  &   0.0327  &   0.0493  &   0.0584  \\ \hline
0.80      &   1.00    &   0.0274  &   \textbf{0.0243}  &   0.0365  &   0.0323  &   0.0508  &   0.0638  \\ \hline
0.90      &   1.00    &   0.0212  &   \textbf{0.0206 } &   0.0363  &   0.0284  &   0.0506  &   0.0621  \\ \hline
\end{tabular}
\end{center}
\def\baselinestretch{1.2}
\caption{\label{table:armax-table} RMSE values for $X_n = \max\{bX_{n-1}, (1-b)Z_n\}$, with $Z_i$
iid standard $1-$Fr\'echet.  The first column contains the $\theta$ values.  The last 6 columns contain the
best RMSE values for the max-spectrum estimates via GLS, WLS, and the competitors.  The sample sizes were fixed at
$2^{13}$, with $N_{out} = 500$, and $N_{in} = 25$.}
\end{table}

\begin{table}[h!]
\scriptsize
\begin{center}
\begin{tabular}{|c|c|c|c|c|c|c|c|}
  \hline
$\theta$  &  $\alpha$ &   $GLS$   & $WLS$     & $F/S$     & $Runs-1$  & $Runs-5$  & $Runs-9$ \\ \hline \hline
0.36    &   0.10    &   0.0226  &   0.0291  &   0.0172  &   \textbf{0.0100}  &   0.0155  &   0.0198    \\ \hline
0.48    &   0.50    &   0.0262  &   0.0299  &   0.0204  &   \textbf{0.0181}  &   0.0322  &   0.0373    \\ \hline
0.63    &   1.00    &   0.0328  &   0.0315  &   \textbf{0.0235}  &   0.0265  &   0.0441  &   0.0509    \\ \hline
0.74    &   1.50    &   0.0226  &   \textbf{0.0203}  &   0.0404  &   0.0333  &   0.0509  &   0.0611    \\ \hline
0.83    &   2.00    &   \textbf{0.0147}  &   0.0238  &   0.0598  &   0.0412  &   0.0576  &   0.0667    \\ \hline
0.89    &   2.50    &   0.0032  &   0.0162  &   0.0007  &   0.0003  &   \textbf{0.0000}  &   0.0007    \\ \hline
0.93    &   3.00    &   0.0013  &   0.0004  &   \textbf{0.0002}  &   0.0043  &   0.0043  &   0.0043    \\ \hline
\end{tabular}
\end{center}
\caption{\label{table:linearprocess-table} RMSE values for $Y_n = 0.50Z_n + 0.20Z_{n-1} + 0.10Z_{n-2}$,
with $Z_i$ iid t-distributed.  The first column contains the $\theta$ values.  The tail index values are in the second
column.  The last 6 columns contain the best RMSE values for the max-spectrum estimates via GLS, WLS, and the
competitors.  The sample sizes were fixed at $2^{14}$, with $N_{out} = 500$, and $N_{in} = 25$.}
\end{table}

\begin{table}[h!]
\scriptsize
\begin{center}
\begin{tabular}{|c|c|c|c|c|c|c|c|}
  \hline
$\theta$  &  $\alpha$ &   $GLS$   & $WLS$     & $F/S$     & $Runs-1$  & $Runs-5$  & $Runs-9$ \\ \hline \hline
0.36    &   0.10    &   0.0212  &   0.0287  &   0.0212  &   \textbf{0.0085 } &   0.0143  &   0.0181    \\ \hline
0.45    &   0.50    &   \textbf{0.0244 } &   0.0311  &   0.0256  &   0.0557  &   0.0274  &   0.0334    \\ \hline
0.57    &   1.00    &   \textbf{0.0315}  &   0.0325  &   0.0329  &   0.0867  &   0.0400  &   0.0474    \\ \hline
0.68    &   1.50    &   0.0353  &   \textbf{0.0340}  &   0.0350  &   0.0844  &   0.0471  &   0.0560    \\ \hline
0.76    &   2.00    &   0.0348  &   \textbf{0.0328}  &   0.0365  &   0.0606  &   0.0482  &   0.0571    \\ \hline
0.83    &   2.50    &   \textbf{0.0320}  &   0.0323  &   0.0378  &   0.0324  &   0.0527  &   0.0625    \\ \hline
0.88    &   3.00    &   0.0301  &   0.0297  &   0.0400  &   \textbf{0.0124}  &   0.0501  &   0.0594    \\ \hline
\end{tabular}
\end{center}
\caption{\label{table:moving-maxima} RMSE values for $W_n = \max\{0.80Z_n, 0.20Z_{n-1}, 0.40Z_{n-2}\}$,
with $Z_i$ iid Pareto.  The first column contains the $\theta$ values.  The tail index values are in the second column.
The last 6 columns contain the best RMSE values for the max-spectrum estimates via GLS, WLS, and the competitors.
The sample sizes were fixed at $2^{13}$, with $N_{out} = 500$, and $N_{in} = 25$.}
\end{table}

\begin{figure}[h!]
\begin{center}
\includegraphics[width=5in]{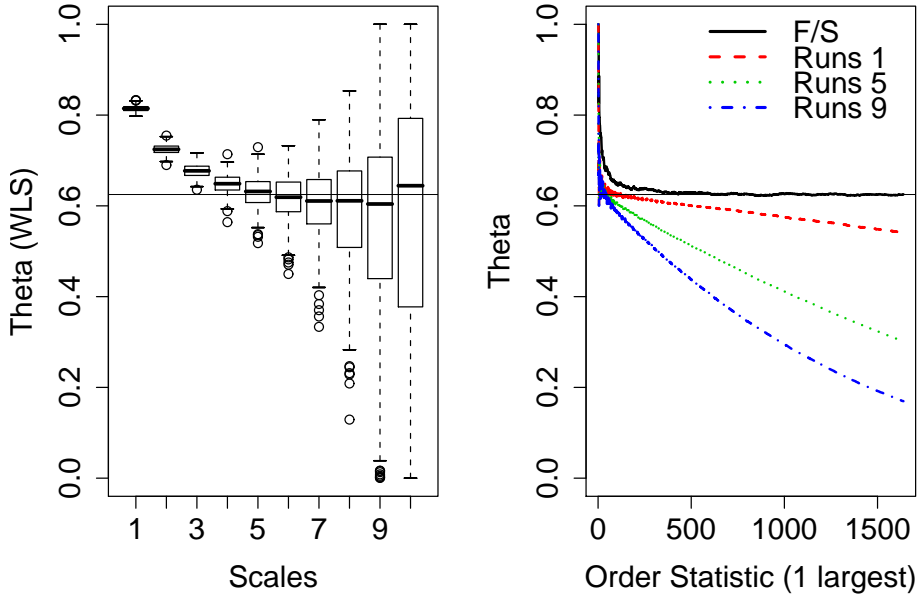}
{\caption{\label{plot:simulationresult} WLS simulation results for
$Y_n = 0.50Z_n + 0.20Z_{n-1} + 0.10Z_{n-2}$, $\theta=0.625$, $Z_i$ iid t-distributed with $df=\alpha=1.00$, and a
sample size of $2^{14}$, with $N_{out}=500$ and $N_{in}=25$.  {\it Left panel:} Boxplots of max-spectrum $\hat{\theta}$.
{\it Right panel:} $\hat{\theta}$ obtained form the runs and Ferro--Segers estimators.  In both plots,
the solid horizontal line corresponds to $\theta=0.625$.}}
\end{center}
\end{figure}

Figure \ref{plot:simulationresult} shows boxplots of $500$ independent  realizations of the WLS variant of the
max--spectrum estimator, computed for a linear process with $\theta=0.625$. The boxplots for the WLS
(GLS boxplots were very similar) method and the median of the estimates of the Ferro--Segers and the runs estimators
per threshold are shown. The runs estimator is quite sensitive to the choice of the run--length and exhibits systematic
bias. The Ferro--Segers and max--spectrum estimators are more robust and do not exhibit such strong bias, a fact
observed in numerous other experimental settings.

\medskip
\noindent{\bf On the choice of $N_{in}$ and $N_{out}$:} The choice of the
 resampling parameters $N_{in}$ and $N_{out}$ in Step 5 of Algorithm 1 involves an intricate bias--variance trade off.
 Our experience with various sample sizes $n$ and values for $N_{in}$ and $N_{out}$ shows that
 larger values for $N_{in}$ lead to smaller variances but larger bias. Extremely large values
 of $N_{out}$ may not yield a good resampling approximation of the distribution of the
 $\hat \theta(j)$'s.
 In real data and/or for smaller samples (e.g.\ up to several thousands),
 we recommend using $N_{in}=1$ and $N_{out} = 200$, for example.  Using $N_{in}=1$ yields
 slightly larger variances, leading to wider confidence intervals, but prevents missing the
 'true value' due to elevated bias.
 For moderate and large samples, and if computation time may be of a lesser concern, we recommend using
 $N_{in}>1$.  The choice of $N_{in}>1$, reduces the variance of the estimators, and as long as the value $N_{in}\times N_{out}$
 is not too large, relative to the available sample size, this does not lead to elevated bias.

\noi{\bf Automatic selection of scales:} We illustrate next the performance of the automatic selection procedure,
introduced in Section \ref{s:implementation}.  We use a subset of the armax, linear and moving maxima
processes, described in the simulation setup above.
As before, for each process, we generate 500 independent realizations, of length $2^{13} = 8192$ for the armax (AM)
and moving maxima (MM) processes and $2^{14} = 16384$ for the linear processes (LP).
We now use $N_{out}=200$ and $N_{in}=1$ and thus we obtain 200 \emph{dependent} estimates
of $\theta$ per scale $j$, for each sample path.  We apply the automatic selection procedure based on the Kruskal--Wallis
test (at a level of $5\%$) for each set of $200$ resampled $\theta$ estimates.  We thus obtain a single $\theta$ estimate
per simulated path.

This procedure is repeated for each independent realization and RMSE values are computed
based on the obtained
$\theta$ estimates from the automatic procedure.  We report the best RMSE value (lowest RMSE value among scales),
the median and the standard deviation of the estimates based on the automatic procedure and the same values corresponding
to the scale at which the best RMSE value was obtained (as in Tables \ref{table:armax-table}--\ref{table:moving-maxima}).

\begin{table}[t!]
 \scriptsize
\begin{center}
\begin{tabular}{|c|c|c||c|c|c||c|c|c|}
  \hline
   \multicolumn{3}{|c}{}            & \multicolumn{3}{|c|}{Best Scale}   & \multicolumn{3}{|c|}{Automatic Selection}\\ \hline \hline
$Process$ & $\theta$  &  $\alpha$ & $RMSE$ & $Median$   & $SD$ & $RMSE$  & $Median$  & $SD$ \\ \hline \hline
$AM$  &   0.20    &   1.00    &   0.0252  &   0.22    &   0.0195  &   0.0439  &   0.22    &   0.0404 \\ \hline
$AM$   &   0.50    &   1.00    &   0.0313  &   0.52    &   0.0268  &   0.0748  &   0.52    &   0.0713 \\ \hline
$AM$   &   0.80    &   1.00    &   0.0257  &   0.81    &   0.0221  &   0.0717  &   0.81    &   0.0702 \\ \hline \hline
$LP$  &   0.48    &   0.50    &   0.0303  &   0.49    &   0.0301  &   0.0672  &   0.48    &   0.0670 \\ \hline
$LP$  &   0.74    &   1.50    &   0.0200  &   0.76    &   0.0154  &   0.0635  &   0.74    &   0.0631 \\ \hline
$LP$  &   0.89    &   2.50    &   0.0230  &   0.87    &   0.0090  &   0.0738  &   0.84    &   0.0620 \\ \hline \hline
$MM$  &   0.45    &   0.50    &   0.0324  &   0.47    &   0.0271  &   0.0513  &   0.47    &   0.0493 \\ \hline
$MM$  &   0.68    &   1.50    &   0.0336  &   0.69    &   0.0276  &   0.0666  &   0.69    &   0.0638 \\ \hline
$MM$  &   0.83    &   2.50    &   0.0337  &   0.85    &   0.0226  &   0.0700  &   0.84    &   0.0686 \\ \hline

\end{tabular}
\end{center}
\caption{\label{table:results-of-auto-selection} Best RMSE values versus the RMSE from the automatic scale
 selection procedure.}

\end{table}

Table \ref{table:results-of-auto-selection} indicates that the automatic selection procedure performs very well
in terms of bias (as compared to the best--RMSE scale).  The RMSE values for the automatic selection method
are larger than the best-scale-RMSE values. This is due to the larger variance as seen from the reported standard
deviations.  Such a behavior is to be expected since the automatic selection procedure does not involve any
knowledge of the true value of $\theta$.  In practice, since $\theta$ is unknown, one cannot identify the best
scale $j$ and hence one cannot achieve the best--RMSE. In such a setting the automatic selection procedure
appears to perform well, by producing estimates with low bias and paying a small price in higher variability.

\noi
{\bf Confidence Intervals:} The following variants of confidence intervals were constructed and compared.
The first, based on asymptotic normality (see Theorem \ref{t:main}), is given by
\begin{equation}\label{e:CI_based_on_normality}
\hat{\theta}(j) \pm z_{(1-q)/2} \hat{\theta}(j)\pi \sqrt{1/6n_j},
\end{equation}
where $z_{(1-q)/2}$ is a $(1-q)/2-$th quantile of the standard normal distribution and
$n$ and $n_j = \lfloor n/2^j \rfloor$  are the total sample size and the number of block--maxima involved in
the calculation of the $Y_j$ statistic, respectively. Table \ref{table:coverage_based_on_normality}
displays coverage probabilities for nominal levels $.05$ and $.10$ for scales $j$ between 4 and 8, where
the $\hat \theta(j)$ estimates typically stabilize. These results are based on $500$ independent realizations
for each process.

\begin{table}[h!]
\scriptsize
\begin{center}
\begin{tabular}{|c|c|c||c|c|c|c|c||c|c|c|c|c|}
  \hline
 \multicolumn{3}{|c|}{}            & \multicolumn{5}{|c|}{90\% - Scales}   & \multicolumn{5}{|c|}{95\%- Scales}\\ \hline \hline
$Process$ &   $\theta$   &   $\alpha$   &   4   &   5   &   6   &   7   &   8   &   4   &   5   &   6   &   7   &   8  \\ \hline \hline
$AM$      &   0.20       &   1.00       &   36  &   72  &   82  &   85  &   89  &   48  &   84  &   90  &   93  &   96 \\ \hline
$AM$      &   0.50       &   1.00       &   88  &   96  &   96  &   96  &   96  &   94  &   99  &   98  &   99  &   99 \\ \hline
$AM$      &   0.80       &   1.00       &   99  &   99  &   99  &   99  &   98  &   100 &   100 &   100 &   100 &   99 \\ \hline \hline
$LP$      &   0.48       &   0.50       &   56  &   81  &   80  &   72  &   65  &   68  &   88  &   85  &   78  &   70 \\ \hline
$LP$      &   0.74       &   1.50       &   94  &   90  &   88  &   84  &   79  &   98  &   95  &   93  &   89  &   83 \\ \hline
$LP$      &   0.89       &   2.50       &   49  &   80  &   90  &   89  &   86  &   62  &   87  &   93  &   93  &   89 \\ \hline \hline
$MM$      &   0.45       &   0.50       &   68  &   95  &   99  &   99  &   99  &   82  &   98  &   100 &   100 &   100 \\ \hline
$MM$      &   0.68       &   1.50       &   93  &   99  &   99  &   100 &   100 &   98  &   100 &   100 &   100 &   100 \\ \hline
$MM$      &   0.83       &   2.50       &   99  &   99  &   100 &   100 &   99  &   100 &   100 &   100 &   100 &   99 \\ \hline

\end{tabular}
\end{center}
\caption{\label{table:coverage_based_on_normality} Coverage probabilities for a selected set of processes using
equation (\ref{e:CI_based_on_normality}).}
\end{table}

\begin{table}[h!]
\scriptsize
\begin{center}
\begin{tabular}{|c|c|c||c|c|c|c|c||c|c|c|c|c|}
  \hline
 \multicolumn{3}{|c|}{}            & \multicolumn{5}{|c|}{90\% - Scales}   & \multicolumn{5}{|c|}{95\%- Scales}\\ \hline \hline
$Process$ &   $\theta$   &   $\alpha$   &   4   &   5   &   6   &   7   &   8   &   4   &   5   &   6   &   7   &   8  \\ \hline \hline
$AM$      &   0.20       &   1.00       &   10  &   33  &   37  &   34  &   31  &   13  &   38  &   43  &   40  &   34 \\ \hline
$AM$      &   0.50       &   1.00       &   34  &   58  &   62  &   61  &   61  &   40  &   66  &   69  &   67  &   68 \\ \hline
$AM$      &   0.80       &   1.00       &   75  &   79  &   79  &   80  &   81  &   83  &   85  &   86  &   88  &   87 \\ \hline \hline
$LP$      &   0.48       &   0.50       &   31  &   61  &   58  &   56  &   53  &   36  &   69  &   66  &   64  &   60 \\ \hline
$LP$      &   0.74       &   1.50       &   79  &   75  &   75  &   71  &   74  &   86  &   82  &   82  &   80  &   80 \\ \hline
$LP$      &   0.89       &   2.50       &   20  &   57  &   75  &   82  &   83  &   28  &   65  &   82  &   90  &   90 \\ \hline \hline
$MM$      &   0.45       &   0.50       &   17  &   55  &   68  &   74  &   79  &   20  &   63  &   75  &   81  &   87 \\ \hline
$MM$      &   0.68       &   1.50       &   31  &   67  &   78  &   81  &   83  &   37  &   79  &   86  &   88  &   90 \\ \hline
$MM$      &   0.83       &   2.50       &   60  &   81  &   84  &   85  &   84  &   70  &   88  &   92  &   91  &   89 \\ \hline

\end{tabular}
\end{center}
\caption{\label{table:coverage_based_on_dependent_sampling} Coverage probabilities for a selected set of processes using
equation (\ref{e:CI_based_on_quantiles}).}
\end{table}

The second type of confidence intervals are based on resampled versions of a single sample path
of the data. The computed  $\theta$ estimates are pooled across a range of
scales with reasonable estimates, and then take the appropriate empirical quantiles:
\begin{equation}\label{e:CI_based_on_quantiles}
 (\hat{\theta}(j_1,j_2)_{(\frac{1-q}{2})},\ \hat{\theta}(j_1, j_2)_{(\frac{1+q}{2})}),
\end{equation}
where $\hat{\theta}(j_1,j_2)_{(\tau)}$ represents the empirical $\tau-$th quantile of the
pooled $\hat \theta(j)$ values across scales $j_1\le j \le j_2$. The coverage probabilities based on
(\ref{e:CI_based_on_quantiles}) are reported in Table \ref{table:coverage_based_on_dependent_sampling}.

Tables \ref{table:coverage_based_on_normality} -- \ref{table:coverage_based_on_dependent_sampling} show coverage
probabilities for the middle range of scales. The confidence intervals based on the asymptotic approximation
tend to over--cover the parameter $\theta$, as compared to the nominal level. On the other hand, the resampled
based confidence intervals tend to undercover $\theta$, on the average. Further, experience shows that
for lower scales, the coverage probabilities suffer substantially due to bias; however, as $j$ increases
the results  rapidly improve.  These results indicate that the information from the two types of confidence
intervals, combined, provides useful ball--park estimates for accurate confidence interval estimates for $\theta$.
The difficult problem of obtaining accurate confidence intervals for $\theta$ which work well in practice will be
the focus of future work.


\section{Applications}
 \label{s:data-analysis}

\noi{\bf Crude Oil Data:} The daily log returns of West Texas Intermediate (WTI) crude oil prices from January 2, 1986 to
March 6, 2007 (5744 observations) are analyzed and the extremal index estimated.
Note that the daily log returns (referred as {\it returns} henceforth) are approximately equal to the daily percentage
changes in the price. WTI represents a benchmark against which all oil bound
for the US is priced at and hence its market is deep and liquid. The data were obtained from
Energy Information Administration (see http://www.eia.doe.gov/).
For a useful reference on oil markets see \cite{geman:2005}.

\begin{figure}[t!]
\begin{center}
\includegraphics[width=5in]{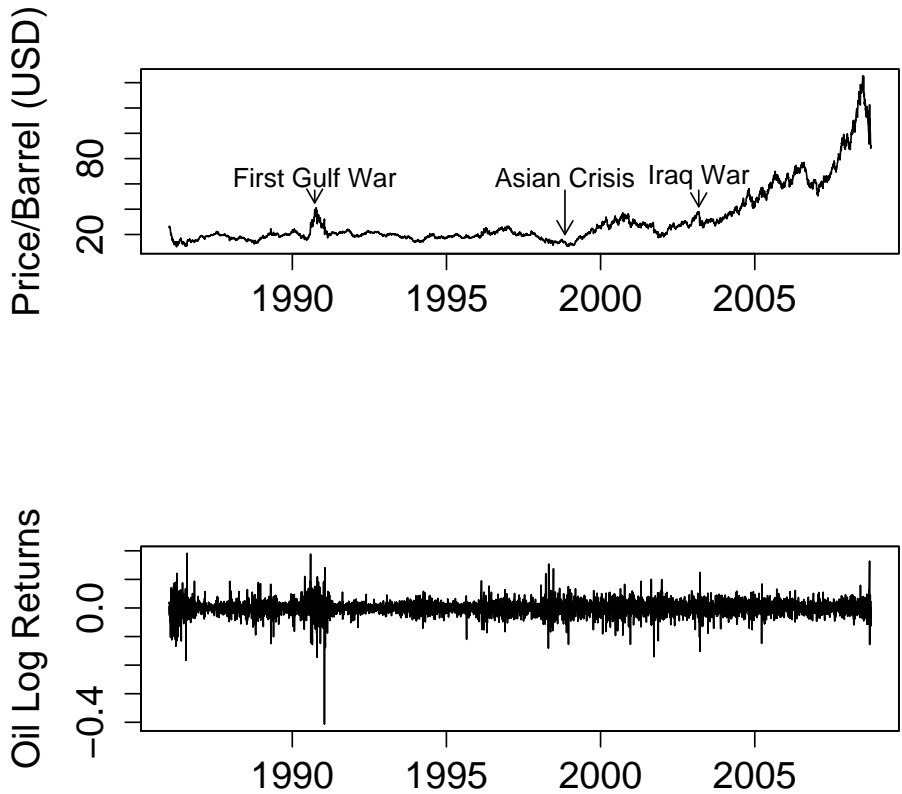}
{\caption{\label{oilplots}  Top Plot: West Texas Intermediate (WTI) crude oil prices from January 2, 1986
to October 7, 2008.  Bottom Plot: The daily log returns of oil prices for the same period. }}
\end{center}
\end{figure}

Figure \ref{oilplots} shows a plot of the data and the corresponding returns. The return series appears to be approximately stationary,
with the exception of a few instances, the result of events of major economic impact.
In the top panel, the run up of the oil prices before the first Persian Gulf war can be seen, together
with its subsequent rapid drop once it became apparent that the coalition forces would prevail.  A similar pattern
is observed at the onset of the recent Iraq war.  The run up in oil prices
over the course of the last three years, accentuated  due to sustained demand and growth is also evident in the plot,
together with their steep drop starting in mid-July 2008.


Analysis of the tail behavior of the data by examining the max-spectrum and Hill estimators
indicate a value of $\alpha \approx 3$ and 2.5 for the right and left tails, respectively.
We study separately the time series of positive (right tail of the distribution) and
negative (left tail) returns. This is motivated by the empirical fact that positive and negative returns
exhibit different behavior.

\begin{figure}[t!]
\begin{center}
\includegraphics[width=5in]{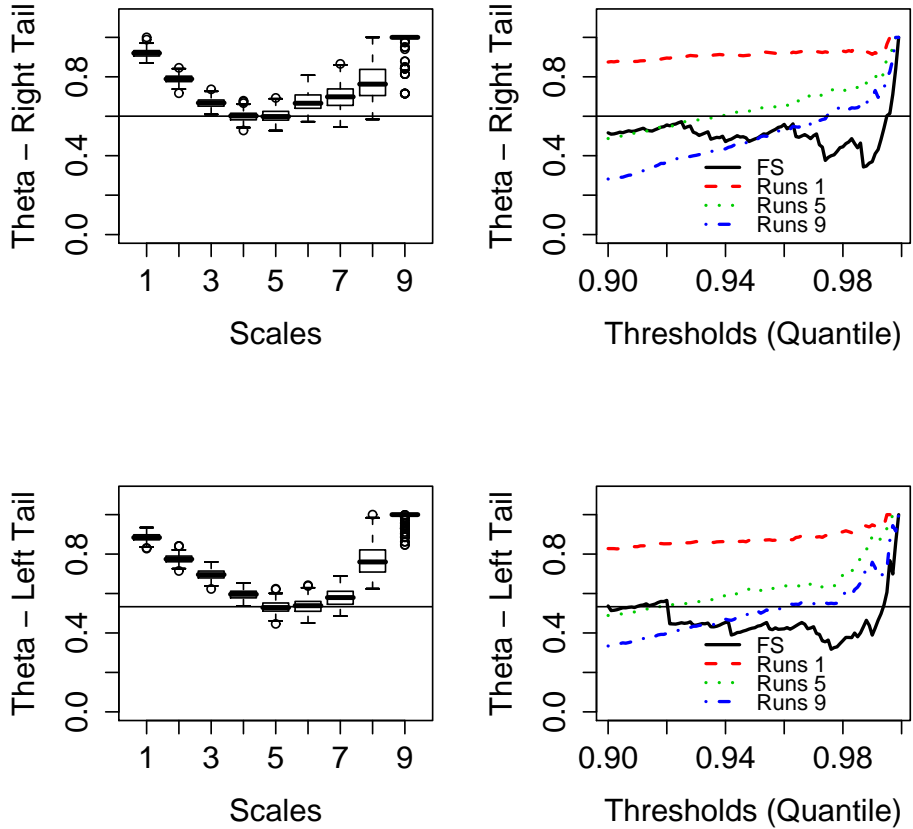}
{\caption{\label{fig:oiltheta}  Top Row: Estimates of $\theta$ for the right tail.  The left panel is the max spectrum estimates.  The right panel is the Ferro--Segers and
runs estimates.  The solid horizontal line in both plots corresponds to the max spectrum point estimate of 0.60.
Bottom Row: Estimates of $\theta$ for the left tail.  The left panel is the max spectrum estimates.  The right panel is the Ferro--Segers and
runs estimates.  The solid horizontal line in both plots corresponds to the max spectrum point estimate of 0.60.}}
\end{center}
\end{figure}

We estimate next the {\it extremal index} $\theta$ of the returns using the max-spectrum, the runs 1, 5, 9 and the Ferro--Segers
estimators. The results are shown in Figure \ref{fig:oiltheta}.
The max-spectrum estimates of $\theta$ were obtained by setting $N_{out}= 200$ and $N_{in}=1$ and using WLS.  It can be seen
that stable $\theta$ estimates for the right tail
can be obtained at scales $j=4$ to $j=5$.  Pooling these results yield a value for $\theta=0.60$
 with a 95\% confidence interval of (0.55, 0.65) based on equation (\ref{e:CI_based_on_quantiles}).
It should be noted that the automatic selection procedure chooses scale $j=5$ for the right tail, which gives
comparable results. The 95\% confidence interval obtained from
 (\ref{e:CI_based_on_normality}) is (0.59, 0.61).
The main reason that these confidence intervals are narrow is because they ignore the
uncertainty regarding scale selection.
For the left tail, we choose the median value at scales $j=5$ to $j=6$ and to obtain a pooled estimate of 0.53
with a 95\% confidence
interval of (0.47, 0.61) using (\ref{e:CI_based_on_quantiles}) and (0.51, 0.55) using
 (\ref{e:CI_based_on_normality}) and $j=5$.

A reasonably stable estimate obtained from the Ferro--Segers procedure is around 0.50 for the
right tail and 0.42 for the left one. However, another choice for the left tail is 0.53, corresponding to
the range of 0.90th to 0.92nd quantiles. The max-spectrum and Ferro--Segers estimates are to some extent
in agreement for the right tail and possibly for the left tail as well, depending on the choice of
a stable range for the Ferro--Segers estimate.
On the other hand, the results of the runs-1 estimator are highly suspect.
The results of the runs-1 indicate little or no clustering of extremes (as $\hat \theta \approx 1$).
The fact that runs-1 fails to capture the clustering may be explained by the behavior of financial returns, where
one extremely large positive return is commonly followed by a large negative return.  Thus, runs-1 often identifies clusters with a
single extreme value, as in the case of independent data.  Increasing the number of the run length parameter yields estimates more
in agreement with the other two procedures. The results strongly suggest clustering of large losses and gains that
can in turn have serious consequences in terms of risk exposure of portfolios that include WTI.

\medskip
\noi The next two examples illustrate our extremal index estimator over two financial data sets: {\it (i)} Daily returns of the
S\& P 500 stock index and {\it (ii)} high--frequency, tick-by-tick volumes of a traded stock.  The extremal index estimates behave
differently in these two settings over the largest scales $j$.  We discuss how the plot of the $\what \theta(j)$'s, as a function
of $j$, may be used to detect different regimes of clustering of extremes. For simplicity, we focus on $\what \theta(j)$'s obtained
by weighted least squares, $N_{in}=1$ and $N_{out} = 200$ independent permutations of the data.  The results with other
choices of the parameters, or ones involving bootstrap instead of permutations are similar.

\begin{figure}[t!]
\begin{center}
\includegraphics[width=5in]{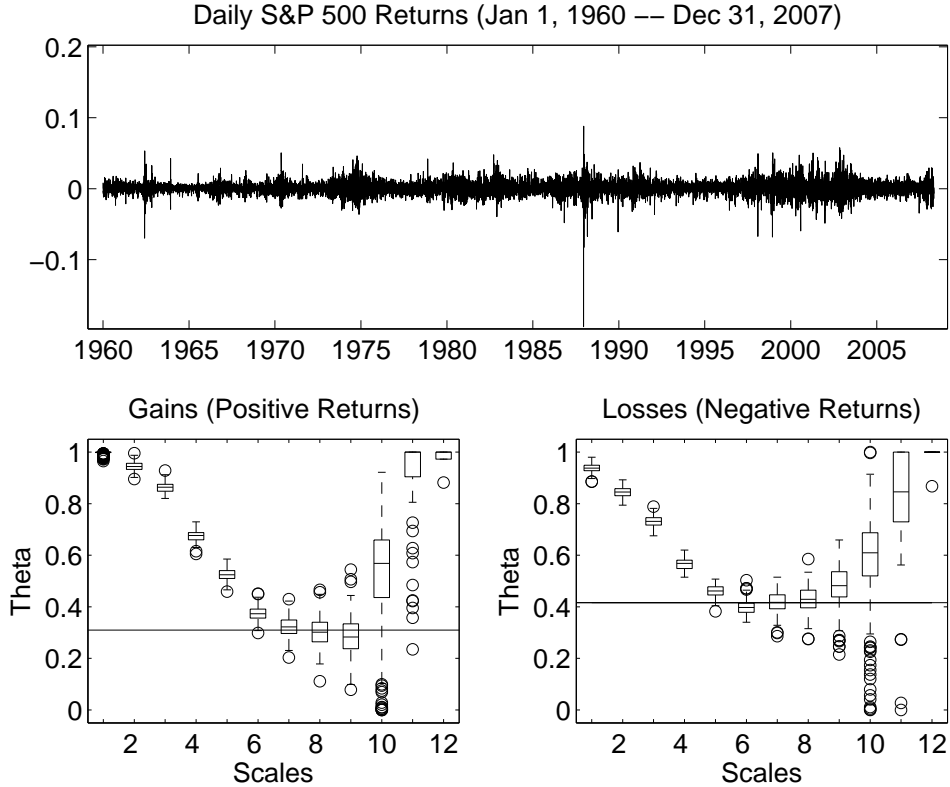}
{\caption{\label{sp500}  {\it Top panel:} S\&P 500 index (daily returns).  {\it Bottom panels:} boxplots of the $\what \theta(j)$'s
obtained from $N_{out}=200$ independent permutations of the data.   (The corresponding bootstrap--based versions are similar and omitted
for brevity.) The {\it left panel} corresponds to the time series of positive returns (gains); the {\it right panel} to the time series
of the absolute values of the negative returns (losses).  Observe that the extremal index estimates over the largest scales approach 1
for both the {\it gains} and the {\it losses}. }}
\end{center}
\end{figure}

\smallskip
\noi{\bf Daily S\&P 500 returns (1960--2007):} Figure \ref{sp500} shows the extremal index estimates of the gains and losses for the
 daily returns of the S\&P 500 stock index.  The top panel indicates that both the gains and the losses time series have heavy tails.
 Indeed, max--spectrum estimates of the left-- and right--tail exponents yield $\what \alpha_{loss} \approx 2.958$ and
 $\what \alpha_{gain} \approx 3.553$.  These values confirm the common observation that the tails of the losses are slightly heavier than
 the tails of the gains (see e.g.\ Table 1 in \cite{galbraith:zernov:2006}).  The bottom two panels on Figure \ref{sp500} show boxplots
 of resampled estimates of the extremal index $\theta$ as a function of the scale $j$.  We studied separately the time series of
 the gains or positive returns (left panel) and the losses (right panel).

 For the {\it gains}, the box--plots stabilize at scales $j=7$ to $9$ (as also confirmed by the Kruskal--Wallis analysis).
 As for the oil data, by pooling the $\what\theta(j)$'s for this range of scales, we obtain $\what \theta_{gains} \approx 0.31$ with
 $95\%$ confidence interval $(0.23, 0.39)$ based on (\ref{e:CI_based_on_normality}) and scale $j = 7$.  The confidence interval based on
 (\ref{e:CI_based_on_quantiles}) and pooling scales $j=7$ to $9$ together is $(0.16,0.43)$.  Similar analysis for the {\it losses}
 shows that the $\what \theta(j)$'s stabilize over the range $j=6$ to $8$, and the pooled estimate is $\what \theta_{loss}
 \approx 0.416$. The  $95\%$ confidence interval based on (\ref{e:CI_based_on_normality}) and scale $j=6$ is $(0.34, 0.49)$, and
 the one based on the pooled scales and (\ref{e:CI_based_on_quantiles}) is $(0.34, 0.50)$.  Our results are in agreement with the
 Ferro--Segers and runs estimates (for 200 threshold exceedences therein) of $\theta_{loss}$ reported in Figure 3b of
 \cite{galbraith:zernov:2006}.

 Our analysis indicates that the extremal indices of both the gains and the losses time series of daily S\&P 500 returns
 are lower than the estimates corresponding to the Oil data set.  This, as before, shows that extremes of the gains and the
 losses exhibit significant clustering, which can have far reaching consequences in terms of risk management.  In contrast to the
 Oil data set, however, the left tails (losses) have slightly higher extremal index than the right tails (gains).  This results in
 slightly more temporal clustering of the extreme gains as compared to the extreme losses.  Indeed, the expected
 cluster sizes for the extreme gains and losses are about $1/\what\theta_{gains} \approx 3.23$ and $1/\what \theta_{loss} \approx 2.5$,
 respectively.

 The above estimates yield a single value for the extremal index $\theta$ based on a judicious choice of scales. In practice, the
 boxplots for the entire range of available scales, however, can also give important insights.  In the above analysis, we focus
 on the range of scales $j=6$ to $9$, which roughly corresponds to focusing on the range of probabilities $[0.9844, 0.9980]$.
 Therefore, from a physical perspective, the extremal index estimates are useful and applicable for the extremes occurring on a
 time scale of up to $2^9 = 512$ trading days or up to 2 years, on the average.  Over a range of 1 to 2 years, one can indeed expect
 that the S\&P 500 returns are approximately stationary and our theory applies.  Significant structural changes and cycles in the economy,
 however, lead to non--stationarity over longer periods of time.  Therefore, the extremal index estimates $\what \theta(j)$'s for scales
 $j \ge 10$ should also be considered, but interpreted with care.  Indeed, as seen from Figure \ref{sp500}, the estimates
 $\what \theta(j)$'s approach 1, as $j$ grows beyond $9$.  For the largest scales ($j=11$ or $12$), the extremal indices of the gains and
 losses are essentially $1$. Since $\theta$ measures the degree of clustering or dependence of extremes, this suggests that the largest
 extremes of the S\&P 500 returns are perhaps weakly dependent or independent.  Indeed, the largest extremes correspond to select few
 financial crashes or periods of extreme volatility.  These events occur far apart in time, they do not cluster, and therefore
 $\what\theta(j) \approx 1$.

\begin{figure}[t!]
\begin{center}
\includegraphics[width=5in]{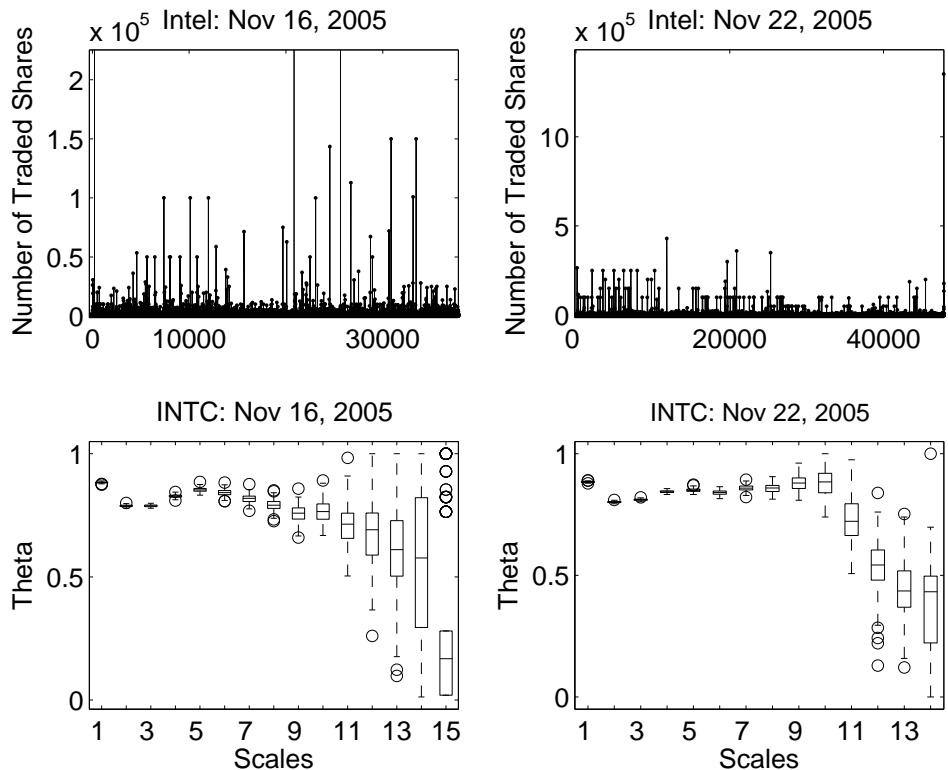}
{\caption{\label{INTC}  {\it Top panels:} High--frequency volume time series for the Intel stock during the days of Nov 16
 (left panel) and Nov 22 (right panel) in 2005.  The data are ordered in time and every value corresponds to the number of traded
 shares during one transaction.  There are 75, 993 trades in Nov 16 and  119, 840 trades in Nov 22.  {\it Bottom panels:} boxplots of the
 $\what \theta(j)$'s obtained from $N_{out}=200$ independent permutations of the data corresponding to the top two data sets.
}}
\end{center}
\end{figure}

\medskip
\noi{\bf High--frequency Stock Volume:}  Figure \ref{INTC} illustrates the extremal index estimators over two high--frequency data
 sets of traded volumes. The time series consist of the number of traded shares of Intel Inc.\ for each and every transaction occurring
 during two typical days of trading (Nov 16 and 22 in 2005).  The data was obtained from the TAQ (trades and quotes) data base of
 consolidated trades from the NYSE and NASDAQ exchanges (see \cite{wharton.data.service}).   One easily sees that reasonable extremal
index estimates for Nov 16 and Nov 22 are about $\what \theta \approx 0.8$.  The corresponding boxplots are stable over a wide range
 of scales (e.g.\ $j=6$ to $10$).  Beyond scales $j=10$, however, one should interpret the estimates $\what \theta(j)$'s with care.
 Indeed, about $2^{10} = 1024$ trades of the highly liquid Intel stock occur over the time scale of about 4 to 5 minutes (depending on
 the time of the day and the amount of trading during the day).  Over the time scale of 4 to 5 minutes, one can
 expect to have relatively stationary trading patterns.  Longer periods of time, however, involve intra--day seasonality and other
 intricate non--stationarity due to arrival of new information.  Therefore, the boxplots of the $\what \theta(j)$'s involve a 'change
 of regime' for scales $j\ge 10$.  This change of regime is relatively abrupt for the November 22 data set and gradual but systematic
 for the November 16 data.  In both cases, the extremal index estimates drastically approach zero, as the scales become more extreme.
 This implies that the clustering of the largest extremes is substantially more pronounced than that of the moderate extremes.  This
effect is also confirmed by the top plots where extremely large volumes are traded in batches.  This phenomenon is in stark contrast with
 the observed weakening of the clustering for the returns data in Figure \ref{sp500}. This difference may be attributed to the difference
 in the nature of the extreme traded volumes and extreme stock returns.  Large returns/losses in the S\&P 500 index are global,
 market--wide events that are hard to control or manipulate.   Extremely large volumes, on the
 other hand, are traded by a select individual players in the market.  Typically, large transactions are split in batches and traded
 systematically over a longer period of time to minimize the negative effect of a large volume trade on the stock price.

\section{Concluding Remarks}

In this paper, a novel procedure for estimating the extremal index of stationary time series was introduced.
It is based on scaling properties of block-maxima and on resampling. Under certain mild regularity conditions, its
consistency and asymptotic normality were established for $m$-dependent time series, that provides one way of
constructing confidence intervals. A more computationally intensive procedure based on resampling is also presented
for the same task. A comprehensive simulation study shows the competitive nature of the proposed estimator. Finally, the
estimator is illustrated on a number of financial data sets that also demonstrate additional diagnostic features
of various statistical plots based on the new estimator.


{\small


\section*{Appendix}

 \label{s:moments}

\begin{prop} \label{p:moments}
Suppose that $f:(0,\infty) \to \mathbb{R}$ is an absolutely continuous function on
any compact interval $[a,b]\subset(0,\infty)$, and such that $f(x) = f(x_0) + \int_{x_0}^x f'(u) du,\ x>0$ 
for some (any) $x_0 >0$.

Let for some $m\in\bbR$ and $\delta>0$,
\begin{equation}\label{e:f-cond-1}
x^m|f(x)| + \esssup_{0<y\le x} y^m |f'(y)| \longrightarrow 0,\ \ \ \mbox{ as }x\downarrow 0,
\end{equation}
\begin{equation}\label{e:f-cond-2}
 x^{-\alpha}|f(x)| + x^{1+\delta} \esssup _{y\ge x} y^{-\alpha} |f'(y)| \longrightarrow 0,\ \ \ \mbox{ as }x\to\infty.
\end{equation}
Suppose also that the time series $X = \{X_n\}_{n\in\bbZ}$ satisfies Conditions 1 and 2,
where $c_1(x)$ is such that:
\begin{equation}\label{e:c1-cond}
 \int_{1}^\infty c_1(x) x^{-\alpha} |f'(x)| dx < \infty.
\end{equation}

\noi Then, $E |f(M_n)| <\infty,$ for all sufficiently large $n\in \mathbb{N}$, and for some $C_f>0$, independent of $n,$
\begin{equation}\label{e:Ef-order}
| \E f(M_n/n^{1/\alpha}) - \E f(Z)| \le C_f n^{-\beta},
\end{equation}
where $Z$ is an $\alpha-$Fr\'echet variable with scale coefficient $\sigma:= c_X^{1/\alpha}$.
\end{prop}

\noi{\bf Proof:} The proof is similar to the proof of Theorem 3.1 in \cite{stoev1}.
Indeed, as in the above reference, one can show that $\E |f(Z)|<\infty $ and $\E |f(M_n)|<\infty$, for all sufficiently large $n$.
Further, by using the conditions \refeq{f-cond-1} and \refeq{f-cond-2} and integration
by parts, we have that
\begin{equation}\label{e:prop:moments-1}
  \E f(M_n/n^{1/\alpha}) - \E f(Z) = \int_0^\infty (G(x) - F_n(x)) f'(x) dx,
\end{equation}
where $F_n(x):= \P\{M_n/n^{1/\alpha} \le x\}$ and $G(x) = \P\{Z\le x\}$. Since $F_n(x) = e^{-c(n,x)x^{-\alpha}},$ by the mean
value theorem, we have
\begin{eqnarray*}
|G(x) - F_n(x)| &=& | e^{- c_X x^{-\alpha}} - e^{-c(n,x) x^{-\alpha}}| \le |c(n,x) -  c_X| x^{-\alpha} e^{-\min\{\theta c_X,\, c(n,x)\} x^{-\alpha}} \\
                &\le& n^{-\beta} c_1(x)x^{-\alpha} {\Big(} e^{-c_2 x^{-(\alpha-\gamma)}} + e^{-\theta c_X x^{-\alpha}} {\Big)},
\end{eqnarray*}
where in the last inequality, we used Relations \refeq{C1} and \refeq{C2}.

Thus, by \refeq{prop:moments-1}, we have that
\begin{eqnarray}\label{e:prop:moments-2}
|\E f(M_n/n^{1/\alpha}) - \E f(Z)| &\le& n^{-\beta} \int_0^\infty c_1(x) x^{-\alpha} |f'(x)| {\Big(} e^{-c_2 x^{-(\alpha-\gamma)}}
             + e^{- c_X x^{-\alpha}} {\Big)} dx \nonumber\\
    &=:& n^{-\beta} {\Big(}\int_0^1 + \int_1^{\infty}{\Big)}.
\end{eqnarray}
The last integral is finite. Indeed, since the exponential terms above are bounded, Relation (\ref{e:c1-cond}) implies that
the integral ``$\int_1^\infty$'' is finite.  On the other hand, conditions \refeq{C1} and \refeq{f-cond-1} imply that,
$c_1(x) |f'(x)| = {\cal O}(x^{-R}),\ x\downarrow 0$, for some $R\in\bbR$.  However, for all $p>0$, we have
 $(e^{-c_2x^{-(\alpha-\gamma)}} + e^{- c_X x^{-\alpha}}) = o(x^{p}),\  x\downarrow 0$, since $\alpha-\gamma>0$.  This implies that
the integral in ``$\int_0^1$'' in (\ref{e:prop:moments-2}) is also finite.  This completes the proof of (\ref{e:Ef-order}).
$\Box$

\medskip
\begin{prop} \label{p:log-moments} Let $X = \{X_k\}_{k \in \mathbb{Z}}$ be a strictly
 stationary time series which satisfies Conditions 1 and 2 in Section \ref{s:theoretical-properties} above. 
 Suppose that $\int_1^\infty c_1(x) x^{-\alpha-1 + \delta} dx <0$, for some $\delta>0$.

 Then, with $M_n := \max_{1\le k\le n} X_k$, we have $\E |\ln(M_n)|^p <\infty$,  for all $p>0$ and all sufficiently large
 $n\in\bbN$. Moreover, for any $p>0$ and $k\in\bbN$, we have:
$$
 {\Big|} \E |\ln(M_n/n^{1/\alpha})|^p - \E |\ln(Z)|^p {\Big|} = {\cal O}(n^{-\beta}),\ \ \mbox{ and }\ \
 {\Big|} \E (\ln(M_n/n^{1/\alpha}))^k - \E (\ln(Z))^k {\Big|} = {\cal O}(n^{-\beta}),
$$
as $n\to\infty$, where $Z$ is an $\alpha-$Fr\'echet random variable with scale coefficient $\theta^{1/\alpha} c_X^{1/\alpha}$.
\end{prop}

\noi{\bf Proof:} It is enough to show that the functions $f(x):= |\ln(x)|^p$ and
 $f(x):= (\ln(x))^k,\ p>0,\ k\in\bbN$ satisfy the conditions of Proposition \ref{p:moments}. In the first case, for example,
 $|f'(x)| = p x^{-1} |\ln(x)|^{p-1},\ x>0$.  Therefore, the assumption $\int_1^\infty c_1(x) x^{-\alpha-1+\delta} dx <\infty $
 implies (\ref{e:c1-cond}), since $ |\ln(x)|^{p-1} \le {\rm const}\, x^{\delta},$ for all $x\in [1,\infty)$.
 The conditions (\ref{e:f-cond-1}) and (\ref{e:f-cond-2}) are also fulfilled in this case, and hence Proposition
 \ref{p:moments} yields the desired order of convergence.  The functions $f(x) = (\ln(x))^k,\ k\in\bbN$ can be
 treated similarly. $\Box$

\medskip
\noi Note that, under the assumptions of Proposition \ref{p:log-moments}, we readily obtain:

\begin{equation}\label{e:E-Yj-rate}
 \E (Y_j - j/\alpha ) \equiv \E \log_2( D(j,k)/2^{j/\alpha} )    = \E \log_2( \theta^{1/\alpha} c_X^{1/\alpha} Z_1 ) + {\cal O}(1/2^{j\beta}),
\end{equation}
as $j\to\infty$, where $Z_1$ is a standard $\alpha-$Fr\'echet variable.
This important fact is used in the proofs of the asymptotic results given below.

\medskip
\noi{\bf Proof of Proposition \ref{p:alpha-C}:} Recall that by (\ref{equ:def_of_BM}),
\begin{equation}\label{e:D-D-tilde}
 D(j,k) := \bigvee_{i=1}^{2^j} X_{2^j(k-1)+i}\ \ \ \ \mbox{ and introduce  }\ \ \ \
 \wtilde{D}(j,k)  :=\bigvee_{i=1}^{2^j-m} X_{2^j(k-1)+i}.
\end{equation}
Observe that $\widetilde{D}(j,k),\ k=1,\ldots, n_j$ ($n_j = \lfloor n/2^j\rfloor $) are independent in $k$ since
they are ``separated by $m$'' block--maxima of the $m-$dependent process $X$.

Recall also that by (\ref{equ:Yj_definition})
$$
Y_j := \frac{1}{n_j} \sum_{k=1}^{n_j} \log_2 D(j,k)\ \ \ \ \mbox{ and introduce the statistics } \ \ \ \
\wtilde Y_j:= \frac{1}{n_j} \sum_{k=1}^{n_j} \log_2 \widetilde{D}(j,k).
$$

We first establish Relation (\ref{e:p:alpha}). Let
\begin{equation}\label{e:H-H-tilde}
\hat{H} = \sum_{i=0}^{\ell} w_i Y_{i+j(n)}, \ \  \ \ \mbox{ and } \ \ \ \ \widetilde{ H } = \sum_{i=0}^{\ell} w_i \widetilde{Y}_{i+j(n)},
\end{equation}
so that $\hat \alpha(j)$  in (\ref{e:alpha-j}) equals $1/\hat H$. The weights $w_i$'s, the range $\ell$ and
the quantity $j(n)$ are described in Section \ref{s:theoretical-properties}.

To prove that $\hat\alpha(j) - \alpha = {\cal O}_P(a_n),\ n\to\infty$, for some $a_n \to 0$, it suffices
to show that $\E (\hat H - H)^2 = {\cal O}(a_n^2)$, where $H:= 1/\alpha$.
Observe that by adding and subtracting the term $\wtilde H$, and by applying the inequality $(a+b)^2 \le 2 a^2 + 2 b^2,$
we get
\begin{eqnarray}
\E (\hat H - H)^2 \le 2 \E (\hat H - \wtilde H)^2 + 2 \E (\wtilde H - H)^2 &=& 2 {\rm Var}(\hat H - \wtilde H)
 + 2 (\E \hat H - \E \wtilde H)^2 + 2 \E (\wtilde H - H)^2\nonumber\\
 &=:& 2A_1 + 2A_2 + 2A_3,\label{e:p:alpha-C-1}
\end{eqnarray}
where in the last relation we also used the fact that  $\E \xi^2 = {\rm Var}(\xi) + (\E \xi)^2$.

\smallskip
{\it We will first show that $A_1 = o(1/n_j)$ in (\ref{e:p:alpha-C-1}) is negligible.}
Indeed, by (\ref{e:H-H-tilde}), we have
\begin{equation}\label{e:H-hat-H-tilde}
 \hat H - \wtilde H = \sum_{i=0}^{\ell} w_i (Y_{i+j(n)} - \wtilde Y_{i+j(n)}),
\end{equation}
and thus by using the inequality ${\rm Var}(\xi_0+\cdots +\xi_{\ell}) \le (\ell+1)^2 ({\rm Var}(\xi_0) + \cdots + {\rm Var}(\xi_\ell))$, we
get
$
{\rm Var}(\hat H - \wtilde H )\le (1+ \ell)^2 \sum_{i=0}^{\ell} w_i^2 {\rm Var}(Y_{i+j(n)} - \wtilde Y_{i+j(n)}).
$
Thus, by Lemma \ref{l:m-dep-1} below, since $\ell$ is fixed,
\begin{equation}\label{e:p:alpha-C-2}
{\rm Var}(\hat H - \wtilde H )\le \frac{{\rm const}}{n_j} \sum_{i=0}^{\ell} {\rm Var}{\Big(}\log_2 D(i+j(n),1) - \log_2 \widetilde{D}(i+j(n),1){\Big)},
\end{equation}
where $n_j= n/2^{j(n)}$. Lemmas \ref{lemma:convergence_of_ratio_of_BM_to_degenrate_RV} and \ref{lemma:uniform_integrable_of_logratios}, on the other
hand, yield
\begin{equation}\label{e:A1}
 {\rm Var}(\hat H - \wtilde H ) = o( 1/n_j),\ \ \ \mbox{ as } n\to\infty.
\end{equation}

\smallskip
{\it Now, we focus on the term $A_2$ in (\ref{e:p:alpha-C-1}).}  By (\ref{e:H-hat-H-tilde}), we have
\begin{eqnarray*}
\sqrt{A_2}  &=& \sum_{i=0}^{\ell} w_i (\E Y_{i+j(n)} - \E \wtilde Y_{i+j(n)}) = \E \sum_{i=0}^{\ell} w_i \log_2(D(i+j(n),1)/2^{(i+j(n))/\alpha}) \\
 & & \ \    -  \E\sum_{i=0}^{\ell} w_i \log_2(\wtilde D(i+j(n),1)/2^{(i+j(n))/\alpha}) \\
  &=& \sum_{i=0}^{\ell} w_i \E \log_2 (Z) + o(1/2^{j(n)\beta}) \\
 & & \ \   - \sum_{i=0}^{\ell} w_i {\Big(}
  \E \log_2( \wtilde D(i+j(n),1)/ (2^{i+j(n)} -m)^{1/\alpha}) - \frac{1}{\alpha} \log_2((2^{i+j(n)}-m)/ 2^{i+j(n)}){\Big)},
\end{eqnarray*}
where the last relation follows from (\ref{e:E-Yj-rate}) and where $Z$ is an $\alpha-$Fr\'echet variable with scale coefficient
$(\theta c_X)^{1/\alpha}$.  Now, since $\wtilde D(i+j(n),1)/(2^{i+j(n)}-m)^{1/\alpha}$ is a properly normalized block--maximum
(recall (\ref{e:D-D-tilde}) above), by Relation (\ref{e:E-Yj-rate}), we further have that
\begin{eqnarray*}
\sqrt{A_2} &= & \sum_{i=0}^{\ell} w_i \E \log_2 (Z) -  \sum_{i=0}^{\ell} w_i \E \log_2 (Z) + o(1/2^{j(n)\beta}) + {\cal O}(\log_2(1 - m/2^{j(n)})\\
 &=& o(1/2^{j(n)\beta}) + {\cal O}(1/2^{j(n)}),
\end{eqnarray*}
as $j(n)\to\infty$, since $\log_2(1-x) = {\cal O}(x),\ x\to 0$.  We thus have,
\begin{equation}\label{e:A2}
 A_2 =  {\cal O}(1/2^{j(n) \min\{1,\beta\}}),\ \ \ \mbox{ as }j(n)\to\infty.
\end{equation}

{\it Consider now the term $A_3$ in (\ref{e:p:alpha-C-1}).} As above, we have
$$
\E (\wtilde H - H)^2 = {\rm Var}(\wtilde H - H) + (\E \wtilde H - H)^2 =: A_3' + A_3'',
$$
and as in (\ref{e:p:alpha-C-2}), we get
$
 A_3' \le (\ell+1)^2\sum_{i=0}^{\ell} w_i {\rm Var}(\wtilde Y_{i+j(n)}) = o(1/n_j) = o(2^{j(n)}/n),\ \mbox{ as } n_j\to \infty.
$
Also, as argued above, since $\sum_{i=0}^{\ell} w_i (i+j(n))/\alpha = 1/\alpha \equiv H$, we obtain
\begin{eqnarray*}
 \E \wtilde H - H 
  = \sum_{i=0}^{\ell} w_i (\E \log_2 \wtilde D(i+j(n),1) - (i+j(n))/\alpha)
 = {\cal O}(1/2^{j(n)\min\{1,\beta\}}),
\end{eqnarray*}
as $j(n)\to\infty$ (see (\ref{e:A2}) above).  By combining the bounds for terms $A_1$, $A_2$ and $A_3$ in (\ref{e:A1}), (\ref{e:A2}) and the last
two relations, we obtain
 $$
  \hat H = H + {\cal O}_P( 1/2^{j(n) \min\{1,\beta\}}) + {\cal O}_P(2^{j(n)/2}/n^{1/2}),\ \ \ \mbox{ as } j(n),\ n/2^{j(n)} \to\infty.
 $$
This completes the proof of the first asymptotic relation in (\ref{e:p:alpha}).

\smallskip
{\it The proof of the second asymptotic relation in (\ref{e:p:C}) is simpler.} By introducing the quantity
$\wtilde C(j): = \wtilde Y_j - j/\alpha$, we have
$$
 C(j) - \wtilde C(j) = Y_j - \wtilde Y_j  = \frac{1}{n_j} \sum_{k=1}^{n_j}\log_2 ( D(j,k)/\wtilde D(j,k)).
$$
One can similarly show that ${\rm Var}(C(j) - \wtilde C(j))$ is of order $o(1/n_j),$ as $n\to\infty$.
Thus, the order of $C(j) - C$ is dictated by the orders of the {\it bias} and {\it standard error} for the
quantity $\wtilde C(j)$.  These can be handled as the terms $A_2$ and $A_3$ in (\ref{e:p:alpha-C-1}).
$\Box$

\medskip
The following three lemmas were used in the proof Proposition \ref{p:alpha-C}.

\begin{lemma} \label{l:m-dep-1} Under the conditions of Proposition \ref{p:alpha-C}, for all $j > \log_2 m$,
we have
$$
 {\rm Var}(Y_j -  \widetilde Y_j ) \le \frac{3}{n_j} {\rm Var}(\log_2 (D(j,1)/ \widetilde{D}(j,1))).
$$
\end{lemma}

\noi{\bf Proof:} For notational simplicity, let $\xi_k:=  \log_2(D(j,k)/ \widetilde{D}(j,k)),\ k=1,\ldots,n_j$.
We have, by the stationarity of $\xi_k$ in $k$, that
$$
{\rm Var}(Y_j - \widetilde{Y_j}) = \frac{1}{n_j} {\rm Var}(\xi_1)  + \frac{2}{n_j^2} \sum_{k=1}^{n_j-1} (n_j-k)
 {\rm Cov}(\xi_{k+1},\xi_1).
$$
Note that $\xi_{k+1} = \log_2(D(j,1+k)/ \widetilde{D}(j,1+k))$ and $\xi_1= \log_2(D(j,1)/ \widetilde{D}(j,1))$
are independent if $k>1$.  Indeed, this follows from the fact that the process $X$ is $m-$dependent, and since $\xi_{k+1}$
and $\xi_1$ depend on blocks of the data separated by at least $2^j > m$ lags.  Therefore, only the
lag--1 covariances in the above sum will be non--zero and hence
$$
{\rm Var}(Y_j - \widetilde{Y_j}) \le \frac{1}{n_j}  {\rm Var}(\xi_1)
+ \frac{2}{n_j} {\Big|}{\rm Cov}(\xi_{2},\xi_1) {\Big|} \le \frac{3}{n_j} {\rm Var}(\xi_1),
$$
since by the Cauchy--Schwartz inequality we have $|{\rm Cov}(\xi_2,\xi_1)| \le {\rm Var}(\xi_2)^{1/2} {\rm Var}(\xi_1)^{1/2} = {\rm Var}(\xi_1)$.
This completes the proof of the lemma.
$\Box$

\begin{lemma} \label{lemma:convergence_of_ratio_of_BM_to_degenrate_RV} For $D(j,k)$ and $\widetilde{D}(j,k)$, 
defined in (\ref{e:D-D-tilde}) above, for any fixed $k$, we have
$ D(j,k) / \widetilde{D}(j,k) \stackrel {P}{\longrightarrow} 1,$ as $j\to\infty.$
\end{lemma}
\noi{\bf Proof:} Let $\delta \in (0,1/\alpha)$ be arbitrary and observe that
\begin{equation}\label{e:D-D-tilde-ineq}
\P\{ D(j,k)/\widetilde D(j,k) < 1 \} = \P \{ R > \widetilde D(j,k)\}
\le \P \{ R > 2^{j\delta} \} + \P\{ 2^{j\delta} > \widetilde D(j,k)\},  \end{equation}
where $R = \max_{1\le i \le m} X_{2^{j}(k-i)+1}$.  Now, by stationarity,
$$\P\{ R > 2^{j\delta}\} = \P\{ \max_{1\le i \le m} X_i > 2^{j\delta}\} \to 0,
\ \ \mbox{ as }j\to\infty.
$$
On the other hand, Relation \refeq{C1} implies that $2^{-j/\alpha} \widetilde D(j,k) \stackrel{d}{\to}
Z,$ as $n\to\infty$, where $Z$ is a non--degenerate $\alpha-$Fr\'echet variable. Thus, since $\delta\in (0,1/\alpha)$,
we have that
$$
\P\{ 2^{j\delta} > \widetilde D(j,k)\} \to 0,\ \ \ \ \mbox{ as } j\to\infty.
$$
The last two convergences and the inequality (\ref{e:D-D-tilde-ineq})
imply that $\P\{ D(j,k)/\wtilde D(j,k) <1\} \to 0,\ j\to\infty$.  Since trivially $\P\{ D(j,k)/\wtilde D(j,k) >1\} =1$, we obtain
$D(j,k)/\wtilde D(j,k)$ converges in distribution to the constant $1$, as $j\to\infty$. 
This completes the proof since convergence in distribution
to a constant implies convergence in probability.
$\Box$

\begin{lemma} \label{lemma:uniform_integrable_of_logratios}
The set of random variables
${\Big|} \log_2 {\Big(} {D(j,k) / \widetilde{D}(j,k)} {\Big)} {\Big|}^p,$ $j, k\in\bbN $
is uniformly integrable, for all $p > 0$, where $D(j,k)$ and $\wtilde D(j,k)$ are defined in (\ref{e:D-D-tilde}).
\end{lemma}

\noi{\bf Proof:}  Let $q>p$ be arbitrary. By using the inequality $|x+y|^q \le 2^{q} (|x|^q + |y|^q),\ \ x,y\in\bbR$,
we get
$$
\E {\Big|}\log_2 \frac{D(j,k)}{ \widetilde{D}(j,k)} {\Big|}^q \le 2^{q} \E |\log_2 (D(j,k)/2^{j/\alpha})|^q  + 2^{q} \E |\log_2 (\wtilde D(j,k)/2^{j/\alpha})|^q.
$$
In view of Proposition \ref{p:log-moments}, applied to the block--maxima $D(j,k)$ and $\wtilde D(j,k)$, we obtain
$$
\E |\log_2 (D(j,k)/2^{j/\alpha})|^q = \E |\log_2(M_{2^j}/2^{j/\alpha})|^q \longrightarrow {\rm const},\ \ \mbox{ as }j\to\infty.
$$
Thus the set $\{\E |\log_2 (D(j,k)/2^{j/\alpha})|^q,\ j,k\in\bbN\}$ is bounded. 
 We similarly have that the set $\{\E |\log_2 (\wtilde D(j,k)/2^{j/\alpha})|^q\}_{j,k\in\bbN}$ is  bounded since $\log_2(2^{j} - m) \sim j,\ j\to \infty$, 
for any fixed $m$.

We have thus shown that
$$\sup_{j,k\in\bbN} \E {\Big|}\log_2 \frac{D(j,k)}{ \widetilde{D}(j,k)} {\Big|}^q <\infty,$$
for  $q>p$, which yields the desired uniform integrability. $\Box$

\medskip
\noi{\bf Proof of Lemma \ref{l:boot-permute}:} Suppose that the indices $i_1,\ldots,i_k$ are drawn without replacement.
Let $A_1 = \Omega$ and
 \begin{equation}\label{e:A_j}
 A_j:=\{\omega\in\Omega\, :\, |i_{j'}(\omega) - i_{j''}(\omega)|\ge m,\ \mbox{ for all } j'\not=j'',\ 1\le j',j''\le j\},
 \end{equation}
 for $j\ge 2$, that is, $A_j$ is the event that the first $j$ random indices are spaced further away from each other
 by at least $m$ lags.  By convention, we let $A_1$ denote the almost sertain event, so that $\P(A_1) = 1$.

We need to show $\P(A_k) \ge 1-mk^2/(n-k)$.  Note that, since $\P(A_1) = 1$ by convention, for all $j\ge 1$, we obtain
\begin{equation}\label{e:boot-1}
 \P(A_{j+1}) = \P(A_{j+1}|A_j) \P(A_j)  \ge  (1-2mj/(n-j))\P(A_j).
\end{equation}
 Indeed, the probability  $\P(A_{j+1}|A_j)$ of choosing the index $i_{j+1}$ to be within $m$ lags from at least one of the chosen
 $j$ indices $i_1,\ldots,i_j$ is at most $2mj/(n-j)$.
 Thus,
 $$
  \P(A_k) = \prod_{j=1}^{k-1} \P(A_{j+1}|A_{j}) \P(A_1) \ge \prod_{j=1}^{k-1} (1- 2mj/(n-j)).
 $$
 Now, by the inequality $\prod_{j=1}^{k-1} (1-x_j) \ge 1 -\sum_{j=1}^{k-1} x_j,$ valid for all $x_j \in [0,1]$, we obtain
 \begin{equation}\label{e:boot-2}
  \P(A_k) \ge 1 -\sum_{j=1}^{k-1} 2mj/(n-j)  \ge 1 - mk(k-1)/(n-k) > 1-mk^2/(n-k).
 \end{equation}
 The case when the indices are drawn with replacement is similar. $\Box$

\noi{\bf Proof of Theorem \ref{t:C-B}:}  Consider either a bootstrap or a permutation
 sample $X^*_l = X_{i_l},\ l=1,\ldots,k$, where $i_1,\ldots,i_k$ are randomly chosen indices from $\{1,\ldots,n\}$,
 independently from the original data $X_1,\ldots,X_n$. In the case of bootstrap these indices are chosen with replacement
 and in the case of permutations -- without replacement, respectively.

 Let the event $A_k$ be defined as in (\ref{e:A_j}), which corresponds to the indices being spaced by at least $m-$lags
 away from each other. Thus, since the time series $X = \{X_i\}_{i\in\bbZ}$ is $m-$dependent,
 $$
 (X_1^*,\cdots, X_k^*)1_{A_k} \stackrel{D}{=} (\wtilde X_1,\cdots, \wtilde X_k) 1_{A_k},
 $$
 where $\wtilde X_l,\ l=1,\ldots,k$ are iid random variables with the same distribution as the $X_n$'s which are
 {\it independent} from the event $A_k$.
 Observe that the event $A_k$ is also independent from the time series $X$ since it depends only on the random indices $i_1,\ldots, i_k$.
 Further, note that in the last relation, we have only {\it equality in distribution} and not equality almost surely.

 Now, by Lemma \ref{l:boot-permute}, we have $\P(A_k) \to 1$, as $k\to\infty$, since $k(n) = o(\sqrt{n})$.
 Thus, Lemma \ref{l:Billingsley} implies that any statistic based on the bootstrap or the randomly permuted
 sample will have the same limiting distribution as the corresponding statistic based on the iid
 sample $\{\wtilde X_{l}\}_{1\le l\le k}$.

 Let $\wtilde C^*(j) = \wtilde Y_{j} - j/\alpha$ be defined as the quantity $C^*(j)$ in (\ref{e:C-j}), but where now
 $\wtilde Y_j$ is the max--spectrum based on the iid data  $\wtilde X_1,\ldots, \wtilde X_k$. Theorem 4.1 in \cite{stoev1} implies that
 \begin{equation}\label{e:wtilde-C}
  \sqrt{k_j} (\wtilde C^*(j) - C^*) \stackrel{D}{\longrightarrow} {\cal N}(0,\sigma_{C^*}^2),\ \ \ \mbox{ as } k\to\infty,
 \end{equation}
 where $\sigma_{C^*}^2$ is as in Theorem \ref{t:C-B}. As argued above, Lemma \ref{l:Billingsley} and Relation (\ref{e:wtilde-C})
 imply (\ref{e:C-B-limit}), which completes the proof of the theorem.  $\Box$

 \begin{lemma}\label{l:Billingsley} Let $X_n,\ X$ and $Y_n$ be real random variables such
  that $X_n \stackrel{D}{\to} X,$ as $n\to\infty$.  Let also
 $A_n$ and $B_n$ be some events such that $Y_n 1_{B_n} \stackrel{D}{=} X_n 1_{A_n}$.
 If $\P(A_n) = \P(B_n) \to 1,\ n\to\infty$, then $Y_n \stackrel{D}{\to} X,$ as $n\to\infty$.
 \end{lemma}
 \noi{\bf Proof:} Let $f:\bbR \to \bbR$ be an arbitrary bounded and continuous function.
 Since $\E |f(Y_n) 1_{B_n^c}| \le {\rm const}\P(B_n^c) = o(1),$ as  $n\to\infty$, we have
 $$
  \E f(Y_n) = \E f(Y_n)1_{B_n} + o(1) = \E f(X_n)1_{A_n} + o(1) = \E f(X_n) + o(1),\ \ \mbox{ as } n\to\infty.
 $$
 This shows that $\lim_{n\to\infty} \E f(Y_n) = \lim_{n\to\infty} \E f(X_n),$ which completes the proof.
 $\Box$

\noi{\bf Proof of Theorem \ref{t:main}:} Recall relation (\ref{e:theta-j}) and observe that by Proposition \ref{p:alpha-C}, we have
$$
 \hat\alpha(j) = \alpha + {\cal O}_P(b_n), \ \ \ \ \mbox{ and } \ \ \ \ C(j) = C + {\cal O}_P(b_n),
 $$
 as $n\to\infty$, where
 $b_n =  1/2^{j(k(n))\min\{1,\beta\}} +  2^{j(k(n))/2}/n^{1/2}.$
%
 Also, by Theorem \ref{t:C-B}, we have
$a_n^{-1}(C^*(j) - C^*) \stackrel{D}{\longrightarrow} {\cal N}(0,\sigma_{C^*}^2),$  as $n\to\infty,$
where $a_n = 1/\sqrt{k_j} = 2^{j(k(n))/2}/k(n)^{1/2}.$
Relation (\ref{e:k(n)-main}), implies that $b_n = o(a_n),\ n\to\infty$. Indeed,
since $k(n) = o(n),\ n\to\infty$, we have $2^{j(k(n))/2}/n^{1/2} = o(2^{j(k(n))}/k(n)^{1/2}) \equiv o(a_n),$
as $n\to\infty$. This shows that the second term of $b_n$ defined above
is negligible with respect to
$a_n$.  By Relation (\ref{e:k(n)-main}), we also have $k/2^{j(k)(1+2\min\{1,\beta\})} \to 0,$ as $k\to\infty$,
or, equivalently $1/2^{j(k)\min\{1,\beta\}} = o(2^{j(k)/2}/k^{1/2}),$ as $k\to\infty$.
Hence, the first term of $b_n$ defined above is also of order $o(2^{j(k(n))/2}/k(n)^{1/2})
\equiv o(a_n)$, as $n\to\infty$.

Now, by using the fact that $b_n = o(a_n),\ n\to\infty$ and the 'Delta--method' 
applied to the function  $f(x,y,z) = 2^{x(y-z)}$ and $x_0=\alpha$, $y_0 = C$ and $z_0 = C^*$
(see also (\ref{e:theta-j})), we obtain
$$
  a_n^{-1}(\hat\theta(j) - \theta)\stackrel{D}{\longrightarrow}
  \partial_z f(\alpha,C,C^*)\, Z \sim {\cal N}(0,\sigma_\theta^2),
$$
as $n\to\infty. $ Since $\partial_zf(x_0,y_0,z_0) = - \ln(2) \alpha \theta $, we obtain
$$
\sigma_\theta^2 = {\Big(}\partial_z f(\alpha,C,C^*) {\Big)}^2 \sigma_{C^*}^2 = \ln(2)^2 \theta^2 {\rm Var}(\log_2(Z)),
$$
where $Z$ is a $1-$Fr\'echet variable (see Theorem \ref{t:C-B}).  Since  $\ln(2) \log_2(Z)$ has the standard Gumbel distribution,
it follows that $\ln(2)^2 {\rm Var}(\log_2(Z)) = \pi^2/6$ (see e.g.\ (22.31) in \cite{johnson_kotz_balakrishnan}).
This completes the proof of the theorem.
$\Box$



\end{document}